\documentclass[10pt,journal,compsoc]{IEEEtran}
\ifCLASSOPTIONcompsoc
  \usepackage[nocompress]{cite}
\else
  \usepackage{cite}
\fi
\ifCLASSINFOpdf
\else
\fi

\usepackage{booktabs} 
\usepackage{subfig}
\usepackage{graphicx}
\usepackage{algorithm}
\usepackage{algpseudocode}%
\usepackage{xcolor}
\usepackage{amsmath, amsthm}
\usepackage{amssymb}
\newcommand{\etal}{\textit{et al}. }
\newcommand{\etals}{\textit{et al}.'s }
\usepackage{mathtools}
\usepackage{multirow}
\usepackage{graphics}
\usepackage{enumitem}
\usepackage{CJKutf8}
\usepackage{array}
\usepackage{diagbox}
\usepackage{svg}
\usepackage{threeparttable}
\usepackage{comment}
\usepackage{xcolor,cite,etoolbox}
\definecolor{mypurple}{rgb}{0.4392, 0.1882, 0.6275}
\definecolor{darkblue}{rgb}{0.0, 0.0, 0.55}
\definecolor{darkgray}{rgb}{0.66, 0.66, 0.66}
\makeatletter 
\pretocmd\@bibitem{\color{black}\csname keycolor#1\endcsname}{}{\fail}
\newcommand \citecolor[1]{\@namedef{keycolor#1}{\color{red}}}
\makeatother

\begin{document}
%
\title{Semi-Supervised Dual-Stream Self-Attentive Adversarial Graph Contrastive Learning for Cross-Subject EEG-based Emotion Recognition}
%
%
%
%

\author{\IEEEauthorblockN{Weishan Ye\textsuperscript{1},
Zhiguo Zhang\textsuperscript{1},
Fei Teng,
Min Zhang,
Jianhong Wang,\\
Dong Ni, 
Fali Li,
Peng Xu\textsuperscript{*}, and
Zhen Liang\textsuperscript{*}}\\
\medskip
\IEEEcompsocitemizethanks{\IEEEcompsocthanksitem Weishan Ye, Fei Teng, Dong Ni, and Zhen Liang are with the School of Biomedical Engineering, Medical School, Shenzhen University, Shenzhen 518060, China, also with the Guangdong Provincial Key Laboratory of Biomedical Measurements and Ultrasound Imaging, Shenzhen 518060, China. E-mail: \{2110246024, 2021222013\}@email.szu.edu.cn, and \{nidong, janezliang\}@szu.edu.cn.
\IEEEcompsocthanksitem Zhiguo Zhang is with the Institute of Computing and Intelligence, Harbin Institute of Technology, Shenzhen 518000, China, and also with the Marshall Laboratory of Biomedical Engineering, Shenzhen 518060, China, and also with the Peng Cheng Laboratory, Shenzhen  518055, China. E-mail: {zhiguozhang@hit.edu.cn.}
\IEEEcompsocthanksitem {Min Zhang is with the Institute of Computing and Intelligence, Harbin Institute of Technology, Shenzhen 518000, China. E-mail: zhangmin2021@hit.edu.cn}.
\IEEEcompsocthanksitem {Jianhong Wang is with Shenzhen Mental Health Center, Shenzhen Kangning Hospital, Shenzhen, China. E-mail: wangjianhong0755@163.com}.
\IEEEcompsocthanksitem Fali Li and Peng Xu are with The Clinical Hospital of Chengdu Brain Science Institute, MOE Key Lab for Neuroinformation, University of Electronic Science and Technology of China, China, and also with School of Life Science and Technology, Center for Information in Medicine, University of Electronic Science and Technology of China, China. E-mail: \{fali.li, xupeng\}@uestc.edu.cn.
\IEEEcompsocthanksitem  \textsuperscript{*}Corresponding author: Peng Xu and Zhen Liang.}}
\IEEEtitleabstractindextext{%
\begin{abstract}
\textcolor{black}{Electroencephalography (EEG) is an objective tool for emotion recognition with promising applications. However, the scarcity of labeled data remains a major challenge in this field, limiting the widespread use of EEG-based emotion recognition. In this paper, a semi-supervised \textbf{D}ual-stream \textbf{S}elf-attentive \textbf{A}dversarial \textbf{G}raph \textbf{C}ontrastive learning framework (termed as \textbf{DS-AGC}) is proposed to tackle the challenge of limited labeled data in cross-subject EEG-based emotion recognition. The DS-AGC framework includes two parallel streams for extracting non-structural and structural EEG features. The non-structural stream incorporates a semi-supervised multi-domain adaptation method to alleviate distribution discrepancy among labeled source domain, unlabeled source domain, and unknown target domain. The structural stream develops a graph contrastive learning method to extract effective graph-based feature representation from multiple EEG channels in a semi-supervised manner. Further, a self-attentive fusion module is developed for feature fusion, sample selection, and emotion recognition, which highlights EEG features more relevant to emotions and data samples in the labeled source domain that are closer to the target domain. \textcolor{black}{Extensive experiments are conducted on three benchmark databases (SEED, SEED-IV and SEED-V) using a semi-supervised cross-subject leave-one-subject-out cross-validation evaluation protocol. The results show that the proposed model outperforms existing methods under different incomplete label conditions with an average improvement of 2.17\%, which demonstrates its effectiveness in addressing the label scarcity problem in cross-subject EEG-based emotion recognition. The source code is available at \textit{https://github.com/Vesan-yws/DS-AGC.}}}
\end{abstract}

\begin{IEEEkeywords}
EEG, emotion recognition, graph contrastive learning, domain adaption, semi-supervised learning.
\end{IEEEkeywords}}

\maketitle

\footnotetext[1]{\hspace{1mm}Equal contributions.}
\IEEEdisplaynontitleabstractindextext

%
\IEEEpeerreviewmaketitle

\IEEEraisesectionheading{
\section{Introduction}
\label{sec:introduction}}
\textcolor{black}{\IEEEPARstart{T}{he} field of affective computing is experiencing rapid growth in Electroencephalography (EEG)-based emotion recognition. However, the predominant focus in current research lies on supervised learning approaches that heavily rely on high-quality labeled data for model training. This process can be labor-intensive, expensive, and complicated to acquire. In contrast, a vast amount of readily available unlabeled data presents an opportunity for semi-supervised learning (SSL), where a small amount of labeled data can be combined with a large amount of unlabeled data to construct models. The generalization capacity could be enhanced, while the burden of extensive labeling efforts could be reduced\cite{van2020survey}.} 

\textcolor{black}{Currently, one of the critical challenges in semi-supervised EEG-based emotion recognition is to develop algorithms that can efficiently utilize both labeled and unlabeled data to improve model learning. For example, Jia \etal \cite{jia2014novel} developed a generative restricted Boltzmann machine (RBM) model and incorporated a regularization of the supervised training process with unlabeled data for model variance reduction. In the model, the unsupervised information was adopted for initial feature selection and treated as learning constraints. In the experiments, the data was randomly divided into two groups: a supervised group and an unsupervised group. This division may result in both groups containing samples from the same subject, potentially affecting the generalization of the model. In a more recent study, Zhang \etal \cite{zhang2022holistic} experimented with a variety of SSL methods on two public emotion EEG databases, demonstrating that effective utilization of unlabeled data can lead to improved model performance. Specifically, they explored three state-of-the-art methods (MixMatch \cite{berthelot2019mixmatch}, FixMatch \cite{sohn2020fixmatch}, and AdaMatch \cite{berthelot2021adamatch}) as well as five classical semi-supervised methods. Their results underscored the importance of leveraging unlabeled data to improve the performance of EEG-based emotion recognition models. However, their evaluation protocol focused exclusively on within-subject assessments. Zhang \etal \cite{zhang2022parse} introduced a domain adaptation method called PARSE that combined domain adversarial neural networks (DANN) and pairwise representation alignment. The results showed that PARSE achieved similar performance to fully-supervised models with substantially fewer labeled samples. Nevertheless, PARSE constrains its utilization to solely labeled data from the source domain for model training, without taking into account the unlabeled data present in the source domain. The previous semi-supervised EEG-based emotion recognition methods have exhibited limitations in their ability to construct robust domain classifiers, which failed to distinguish among the labeled source domain, the unlabeled source domain, and the unknown target domain. This deficiency has led to suboptimal adaptation to the unlabeled source domain, especially in scenarios where the source data predominantly consists of unlabeled samples. Consequently, these methods have not achieved the desired level of performance in such challenging situations. Additionally, these methods extracted features from isolated EEG channels, which disregard the complex feature representation among different EEG channels. This limitation results in a lack of rich structural information representation in modeling.} 

\textcolor{black}{To address the limitations in the literature, this paper proposes a novel semi-supervised \textbf{D}ual-stream \textbf{S}elf-attentive \textbf{A}dversarial \textbf{G}raph \textbf{C}ontrastive learning framework (termed as \textbf{DS-AGC}) for cross-subject EEG-based emotion recognition. A dual-stream SSL framework is designed, comprising a non-structural stream and a structural stream. Within this framework, two types of feature representations are generated, ensuring the preservation of both non-structural and structural feature information.} In the non-structural stream, differential entropy (DE) features are extracted from labeled source data, unlabeled source data, and unknown target data, and a semi-supervised multi-domain adaptation method is developed to jointly minimize distribution discrepancy among the three domains. This approach helps to address the limitations of previous methods by leveraging unlabeled source data to improve model performance. In the structural stream, a graph contrastive learning method is proposed to extract effective graph-based feature representation from multiple EEG channels in a semi-supervised manner. This approach can capture the complex relationships between features across different channels and address the limitation of previous methods in overlooking structural information. Finally, to efficiently and effectively fuse the dual-stream feature representation in a delicate and intelligent manner, a self-attentive fusion module is introduced, leveraging the strengths of both streams and highlighting the EEG features and samples that are more discriminative for emotions. Overall, the proposed DS-AGC can overcome the limitations of previous methods in adapting to the unlabeled source domain and capturing rich non-structural and structural information. \textcolor{black}{The main contributions of this study can be summarized as follows.}

\begin{itemize}
\item \textcolor{black}{We propose a novel semi-supervised learning framework designed to efficiently leverage a limited set of labeled data in conjunction with a large quantity of unlabeled data, ultimately enhancing the performance of semi-supervised cross-subject EEG emotion recognition.}
\item \textcolor{black}{A dual-stream architecture is designed to capture and extract a comprehensive array of EEG features, including both non-structural and structural elements. Both streams are intelligently united through a self-attentive module, ensuring the preservation of the intricate blend of non-structural and structural information inherent within EEG signals.}
\item \textcolor{black}{Extensive experiments are conducted on benchmark databases, covering various data scarcity conditions. The effectiveness and reliability of the model's performance are validated in a cross-subject scenario, while thorough analyses of model components and parameters, along with feature visualization, are performed to enhance our understanding of the results.}

\end{itemize}

\section{Related Work}
\label{sec:relatedWork}
\subsection{EEG-based emotion recognition}
In recent years, there has been an increasing interest in the potential of EEG signals to identify emotional states of individuals{\cite{De2013emotion,zheng2015investigating,li2016eeg,liu2017real,zhang2018cascade,yang2019multi,khare2020adaptive,anuragi2022eeg}. These studies have yielded valuable insights and paved the way for the development of affective brain-computer interface (aBCI) systems. Among the early contributions to this field, Duan \etal extracted emotion-related EEG features and employed support vector machines (SVM) for emotion classification\cite{De2013emotion}. \textcolor{black}{Since then, a number of studies based on various machine learning and deep learning methods have been developed to enhance the accuracy and usability of aBCI systems.}

The recent advancements in deep learning have led to a surge of research on EEG-based emotion recognition using various neural network architectures. For example, Zhang \etal proposed a cascaded and parallel convolutional recursive neural network that can effectively learn discriminative EEG features\cite{zhang2018cascade}. Song \etal introduced a dynamic graph convolutional neural network to dynamically capture intrinsic relationships among different EEG channels\cite{song2018eeg}. Niu \etal developed a novel deep residual neural network by combining brain network analysis and channel-spatial attention mechanism\cite{niu2023brain}. The above-mentioned models were robust in within-subject emotion recognition tasks, where the training and testing data come from the same subject. However, due to individual differences in EEG signals collected from different subjects, the model performance would significantly decrease in cross-subject emotion recognition tasks, when the training and testing data come from different subjects\cite{zheng2016personalizing,samek2013transferring,morioka2015learning}.

\subsection{Cross-subject EEG-based emotion recognition}
\textcolor{black}{To improve the generalizability of cross-subject EEG-based emotion recognition models, transfer learning is employed to mitigate individual differences by harmonizing the feature distribution across different subjects \cite{xu2022dagam, huang2022generator, he2022adversarial, zheng2016personalizing, jin2017eeg}.} 
Transfer learning methods, such as transfer component analysis (TCA) and transductive parameter transfer (TPT), have been incorporated in Zheng \etals work\cite{zheng2016personalizing} to improve cross-subject emotion recognition performance from 56.73\% to 76.31\%, demonstrating the effectiveness of transfer learning in improving the generalizability of EEG-based emotion recognition models in the presence of individual differences among subjects.

In recent years, deep transfer learning methods have been proposed to further improve the model performance in cross-subject EEG-based emotion recognition tasks. In 2017, Jin \etal\cite{jin2017eeg} introduced a deep transfer learning framework with DANN\cite{ganin2016domain} to further improve the emotion recognition accuracy from 76.31\% (non-deep transfer learning) to 79.19\% (deep transfer learning). \textcolor{black}{Based on Jin \etals work, Li \etal proposed a domain adaptive method to minimize distribution shift and generalize the emotion recognition model to different individuals\cite{li2019domain}.} He \etal combined adversarial discriminators and time convolutional networks (TCNs) to further enhance distribution matching in EEG-based emotion recognition tasks\cite{he2022adversarial}. To preserve the neural information during feature alignment, Huang \etal developed a generator-based ignorant mechanism domain adaptation model \cite{huang2022generator}. Xu \etal proposed a domain adversarial graph attention model (DAGAM) that utilized graph attention adversarial training and biological topology information\cite{xu2022dagam}. Additionally, Peng \etal proposed a joint feature adaptation and graph-adaptive label propagation method (JAGP)\cite{peng2022joint}. Compared to the original DANN structure, these enhanced deep transfer learning frameworks have achieved better performance on the cross-subject EEG-based emotion recognition tasks.

\textcolor{black}{Recent deep transfer learning-based emotion recognition models have shown promising results in handling individual differences. Nonetheless, these models often necessitate a significant volume of labeled source data to establish consistent performance. Acquiring a sufficient amount of high-quality labeled EEG data is a challenging and time-consuming task in aBCI systems. In light of this challenge, the development of a novel SSL framework for cross-subject EEG-based emotion recognition becomes important. Such a framework should effectively leverage a small quantity of labeled source data and a large amount of unlabeled source data. This represents a crucial research direction with the potential to mitigate the constraints imposed by the limited availability of labeled data in EEG-based emotion recognition.}

\subsection{Contrastive learning}
\textcolor{black}{Contrastive learning offers a promising alternative for feature learning, as it enables the extraction of meaningful representations directly from data without the need for manual annotation.} This approach has found widespread application in various fields, including computer vision \cite{chen2020simple}, natural language processing (NLP) \cite{devlin2018bert}, and bioinformatics \cite{li2021effective,liu2021deep}. In recent years, researchers have also started exploring the potential of contrastive learning for EEG-based emotion recognition. For example, Mohsenvand \etal \cite{mohsenvand2020contrastive} utilized the SimCLR framework \cite{chen2020simple} to learn the similarity between augmented EEG samples from the same input, thereby enhancing the model's ability to capture distinctive features. Similarly, Shen \etal \cite{shen2022contrastive} proposed the CLISA model, which maximized feature representation similarity across subjects experiencing the same emotional stimuli, leading to improved emotion recognition performance. These studies highlight the efficacy of contrastive learning in the context of EEG-based emotion recognition. \textcolor{black}{However, there are still challenges that need to be addressed in contrastive learning based EEG modeling, particularly with regard to effectively integrating structural information from EEG signals into the contrastive learning framework.}

\textcolor{black}{Exploring the structural information present in EEG signals is vital for capturing the complex connectivity patterns and topographical relationships within the brain. EEG signals exhibit natural spatial and temporal dependencies among different electrodes and brain regions, which can provide valuable insights into brain dynamics and emotional processes. Graph contrastive learning (GCL) has emerged as a powerful technique that leverages structured information in data, expanding the capabilities of contrastive learning to incorporate rich contextual information in graph data \cite{you2020graph}. Integrating GCL into EEG analysis holds the potential for facilitating a more comprehensive analysis of brain function and its implications, specifically by considering the interplay between different EEG channels within the complex and dynamic realm of emotional processing. The intricate connectivity patterns within the brain could be uncovered and the understanding of how different channels interact and contribute to the overall emotional experience could be deepened.}


\section{Methodology}
\label{sec:methodology}

\begin{figure*}
\begin{center}
\includegraphics[width=\textwidth]{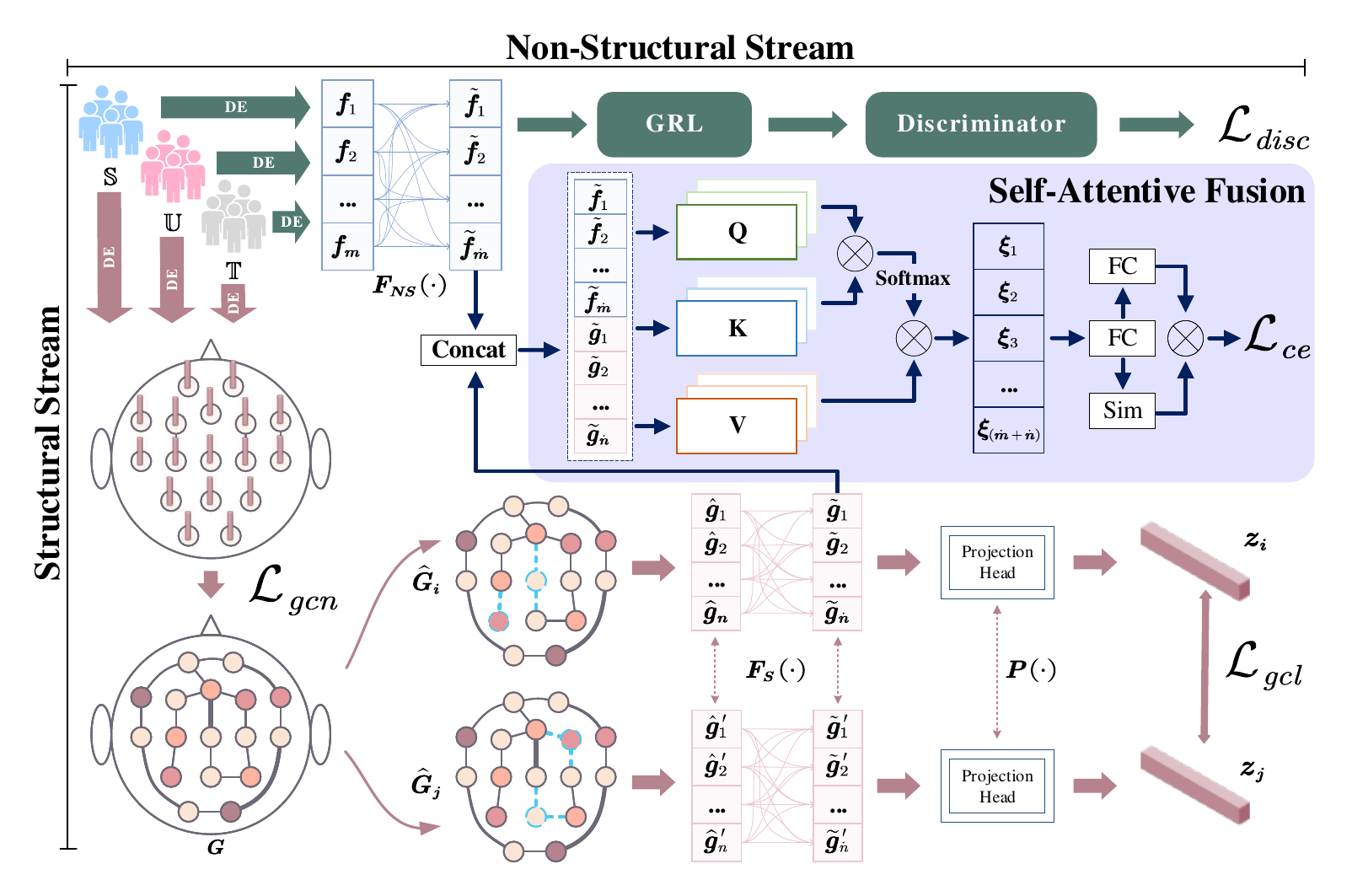}
\end{center}
\caption{\textcolor{black}{An overview of the proposed DS-AGC. It consists of three parts: non-structural stream, structural stream, and self-attentive fusion. $\mathbb{S}$, $\mathbb{U}$, and $\mathbb{T}$ refer to the labeled source domain, unlabeled source domain, and unknown target domain.}}
\label{fig:standard_pipeline}
\end{figure*}

\textcolor{black}{An overview of the proposed DS-AGC is shown in Fig. \ref{fig:standard_pipeline}, which consists of three main parts. \textbf{(1) Non-Structural Stream:} extracting non-structural EEG features from the labeled source $\mathbb{S}$, unlabeled source $\mathbb{U}$, and unknown target $\mathbb{T}$ domains. Based on the extracted DE features at five different frequency bands, a non-structural feature extractor $F_{NS}(\cdot)$ is defined to extract the features from each domain. A gradient reversal layer is used to reverse the gradients during the feature extraction process, allowing the model to learn domain-invariant features and ensuring that the extracted features from the three domains are indistinguishable. \textbf{(2) Structural Stream:} extracting structural EEG features from $\mathbb{S}$, $\mathbb{U}$, and $\mathbb{T}$. Based on the extracted DE features at five different frequency bands, a graph convolution network (GCN) is constructed for spatial feature representation, and the corresponding positive samples are generated by data augmentation. Then, a structural feature extractor $F_S(\cdot)$ is defined to characterize structural feature representation from the input, with a contrastive loss ensuring that the structural features extracted from the positive samples are consistent with each other. \textbf{(3) Self-Attentive Fusion:} fusing the extracted non-structural and structural features from the above two parallel streams. A concatenation of the extracted non-structural and structural features is fed into a multi-head self-attention mechanism. This fusion process generates a new feature representation that emphasizes the most discriminative features related to emotions and suppresses irrelevant information. To ensure the decodability of the features, we train a classifier on the labeled source data ($\mathbb{S}$) using the fused features, with a classification loss. The feature representation and the classifier are optimized jointly, ensuring the final feature representation is effective enough for emotion recognition.}

\subsection{Non-Structural Stream}
\label{sec:non-structrual}

\textcolor{black}{Traditional EEG-based supervised domain adaptation methods primarily utilize DANN \cite{ganin2016domain}, aiming to align feature distributions between source and target domains \cite{Du2020,2018WGAN,BiHDM2019,He2018DAN,Zhaoplug2021,zhong2020eeg}. However, in semi-supervised learning, treating labeled and unlabeled source data as a single domain complicates adaptation and adversely affects performance of downstream tasks \cite{li2019multisource,chen2021msmda}.}
\textcolor{black}{To address distribution mismatches among different domains, this paper proposes a novel semi-supervised multi-domain adaptation method, which aligns feature representations among the labeled source domain $\mathbb{S}$, the unlabeled source domain $\mathbb{U}$, and the target domain $\mathbb{T}$. The feature distribution discrepancies across these domains could be mitigated and the model's generalization capabilities could be further enhanced.}

In the proposed model, three domains are defined below. The labeled source domain $\mathbb{S}$ ($D_{s}=\{X_s,Y_s\}=\{(x_{i}^s,y_{i}^s)\}_{i=1}^{N_s}$), which contains labeled samples with their corresponding emotion labels; the unlabeled source domain $\mathbb{U}$ ($D_{u}=\{X_u\}=\{(x_{i}^u)\}_{i=1}^{N_u}$), which contains unlabeled samples without emotion labels; and the unknown target domain $\mathbb{T}$ ($D_{t}=\{X_t\}=\{(x_{i}^t)\}_{i=1}^{N_t}$), which contains samples from an unseen domain that needs to be classified. Here, $x_{i}^s$, $x_{i}^u$, and $x_{i}^t$ are the EEG data from the three domains, and $y_{i}^s$ is the given emotion label of $x_{i}^s$ in the labeled source domain. $N_s$, $N_u$, and $N_t$ are the corresponding sample sizes. It is noted that the emotion label information in the unlabeled source domain and the unknown target domain are not available during model training.
\textcolor{black}{Finally, the three domains, $\mathbb{S}$, $\mathbb{U}$, and $\mathbb{T}$, constitute a multi-domain.}

\textcolor{black}{For the existing domain adaptation approaches involving two domains (the labeled source domain $\mathbb{S}$ and the target domain $\mathbb{T}$), the corresponding target error $\epsilon_T\left(h\right)$ is constrained by the following inequality \cite{ben2010theory}}
\begin{equation}
\color{black}
\label{Eq:targetbound}
\epsilon_{T}(h) \leq \delta+\epsilon_{S}(h)+d_{\mathcal{H} }\left(\mathcal{D}_{S}, \mathcal{D}_{T}\right),
\end{equation}
\textcolor{black}{where $\delta$ represents the difference in labeling functions across the two domains, typically assumed to be small under the covariate shift assumption \cite{david2010impossibility}. $\epsilon_S\left(h\right)$ denotes the source error for a classification hypothesis $h$ as determined by the source classifier. $d_\mathcal{H}\left(\mathcal{D}_S,\mathcal{D}_T\right)$ measures the divergence between the source domain distribution $\mathcal{D}_S$ and the target domain distribution $\mathcal{D}_T$, estimated directly from the error of a trained binary classifier \cite{ben2006analysis}. In the semi-supervised learning, accurately estimating source error and $\mathcal{H}$-divergence with limited labeled source data is highly challenging. To address this, our method utilizes both labeled and unlabeled source samples for estimating source error and $\mathcal{H}$-divergence. The source domain is defined as $\mathbb{S}^\ast=\{\mathbb{S},\mathbb{U}\}$, combining labeled source data ($\mathbb{S}$) and unlabeled source data ($\mathbb{U}$). The convex hull $\mathrm{\Lambda}_{S_\pi^\ast}$ of $S_\pi^\ast$ is a set of mixed distributions, defined as}
\begin{equation}
\color{black}
    \Lambda_{S_{\pi}^*}=\{\overline{\mathcal{D}}_{S_{\pi}^*}: \overline{\mathcal{D}}_{S_{\pi}^*}(\cdot)=\pi_{S}\mathcal{D}_{S}(\cdot)+\pi_{U}\mathcal{D}_{U}(\cdot)\},
\end{equation}
\textcolor{black}{where ${\overline{\mathcal{D}}}_{S_\pi^\ast}$ is a distribution calculated by the weighted sum of the labeled source domain distribution $\mathcal{D}_S$ and the unlabeled source domain distribution $\mathcal{D}_U$. $\pi_S$ and $\pi_U$ are the respective weights, belonging to the simplex $\mathrm{\Delta}_1$. For the target domain $\mathbb{T}$, ${\overline{\mathcal{D}}}_T$ is the nearest element to $\mathcal{D}_T$ within $\mathrm{\Lambda}_{S_\pi^\ast}$, given as}
\begin{equation}
\color{black}
    \operatorname{argmin}_{\pi_{S},\pi_{U}} d_{\mathcal{H}}\left[\mathcal{D}_{T},\pi_{S} \mathcal{D}_{S}+\pi_{U} \mathcal{D}_{U}\right].
\end{equation}
\textcolor{black}{Then, based on prior domain adaptation methods \cite{albuquerque2019generalizing,ben2006analysis,zhao2018adversarial}, a generalization upper-bound for the target error $\epsilon_T\left(h\right)$ can be derived as}
\begin{equation}
\color{black}
\label{Eq:targetbound_triple}
\begin{split}
\epsilon_{T}(h) &\leq \epsilon_{S}(h)+\min \left\{\mathbb{E}_{\overline{\mathcal{D}}_{T}}\left[\left|F_{\overline{T}}-F_{T}\right|\right], \mathbb{E}_{\mathcal{D}_{T}}\left[\left|F_{T}-F_{\overline{T}}\right|\right]\right\}\\&+\pi_{U}(d_{\mathcal{H}}\left(\mathcal{D}_{S}, \mathcal{D}_{U}\right)+d_{\mathcal{H}}\left(\mathcal{D}_{U}, \mathcal{D}_{T}\right))+\pi_{S}d_{\mathcal{H}}\left(\mathcal{D}_{S}, \mathcal{D}_{T}\right)\\&+\pi_U\min \left\{\mathbb{E}_{\mathcal{D}_{S}}\left[\left|F_{S}-F_{U}\right|\right], \mathbb{E}_{\mathcal{D}_{U}}\left[\left|F_{U}-F_{S}\right|\right]\right\},
\end{split}
\end{equation}
\textcolor{black}{where $\epsilon_S\left(h\right)$ is the error of the labeled source domain, and $d_\mathcal{H}\left(\cdot\right)$ denotes the $\mathcal{H}$-divergence between the specified domains. $F_{\bar{T}}\left(x\right)=\pi_SF_S\left(x\right)+\pi_UF_U\left(x\right)$ is the labeling function for any $x\in\mathrm{Supp}\left({\bar{\mathcal{D}}}_T\right)$, representing the target domain's labeling function. In extreme cases where all samples in the source domain are labeled (i.e., $\mathbb{U}$ is empty), the upper bound in Eq. \ref{Eq:targetbound_triple} equals the bound given in the traditional supervised domain adaptation method. Next, we will demonstrate how to empirically optimize the target error in the downstream tasks.}

\textcolor{black}{For the non-structural stream,} we focus on extracting non-structural EEG features. To address the distribution shift among the three domains, a multi-domain adversarial neural network is incorporated for feature adaptation. It aligns the feature distributions, making them more consistent and reliable across the three different domains. Specifically, we first flatten the extracted DE features into one-dimensional feature vectors (termed as $\{f_1,f_2,...,f_m\}$ in Fig. \ref{fig:standard_pipeline}, $m$ is the feature dimensionality) and input them into a feature extractor $F_{NS}(\cdot)$ for sample feature extraction. This produces the corresponding features as $\{\tilde{f}_1,\tilde{f}_2,...,\tilde{f}_{\dot{m}}\}$, where $\dot{m}$ is the obtained feature dimensionality after $F_{NS}(\cdot)$. To align the distribution shift among the extracted $F_{NS}(X_s)$, $F_{NS}(X_u)$, and $F_{NS}(X_t)$ for the labeled source domain, unlabeled source domain, and unknown target domain, we introduce a discriminator $d(\cdot)$ with parameters $\theta_{d}$ to distinguish the domain from which the sample features originate. \textcolor{black}{The distribution discrepancies among the three domains are minimized by optimizing the discriminator loss function. For an ideal joint hypothesis across the \(\mathbb{S}\), \(\mathbb{U}\), and \(\mathbb{T}\) domains, the second term \(\min\{\mathbb{E}_{\overline{\mathcal{D}}_T}[|F_{\overline{T}}-F_T|], \mathbb{E}_{\mathcal{D}_T}[|F_T-F_{\overline{T}}|]\}\) and the last term \(\pi_U \min \{\mathbb{E}_{\mathcal{D}_S}[|F_S-F_U|], \mathbb{E}_{\mathcal{D}_U}[|F_U-F_S|]\}\) in Eq. \ref{Eq:targetbound_triple} are assumed to be small under the covariate shift assumption \cite{david2010impossibility,ben2010theory,albuquerque2019generalizing,sicilia2021domain}. }

\textcolor{black}{Previous research has demonstrated that minimizing the \(\mathcal{H}\)-divergence can be approximated by maximizing the classification error of the domain discriminator through adversarial training \cite{ganin2016domain,ben2006analysis}. Given the distribution differences among the three domains, the domain discriminator loss is redefined in this study as}
\begin{equation}
\label{Eq:Tripledomainloss}
\begin{aligned}
\mathcal{L}_{d i s c}\left(\theta_{f}, \theta_{d}\right)=-\sum_{x_i} l\left(x_i\right) \log d\left(F_{NS}\left(x_i\right)\right),
\end{aligned}
\end{equation}
where $l\left(x_i\right)$ is a one-hot domain label of the input sample data $x_i$, and $F_{NS}(x_i)$ is the corresponding extracted non-structural features. The domain adversarial training aligns the feature representation distribution across the three domains, making the non-structural features more robust to domain changes and more effective for downstream tasks.

\subsection{Structural Stream}
\label{sec:structural}
\textcolor{black}{We capture the structural information of EEG signals collected from multiple EEG channels and express the inherent relationships among the channels by extracting the structural features. It enables us to gain more valuable insights into the complex interconnections and dependencies within the EEG network.} An undirected graph $G=(V^G, E^G, A^G)$ is defined, where $V^G$, $E^G$, ${A^G}$ represent the nodes, edges, and adjacency matrix, respectively. $|V^G|$ is equal to the number of the EEG channels, given as $N_G$. For a sample data $x_i$, the input is denoted as $\boldsymbol{\Psi}_{i}=\left(\boldsymbol{\psi}_{1}^{i}, \boldsymbol{\psi}_{2}^{i}, \cdots, \boldsymbol{\psi}_{N_G}^{i}\right)^{T} \in \mathbb{R}^{N_G\times C_{de}}$, where $\boldsymbol{\psi}_{k}^{i} \in \mathbb{R}^{C_{de}} (k \in{1,2,\cdots,N_G})$ is the extracted DE features of the $k$th node.

\subsubsection{Graph convolution network}
\label{sec:GCN}
A GCN is developed to aggregate the neighbor information of the feature matrix $\boldsymbol{\Psi}_{i}$ for spatial feature extraction, resulting in $G$. Specifically, we construct a channel-based graph, where the nodes $V^G$ correspond to EEG channels, and the node features are the extracted DE features at each channel termed as $\boldsymbol{\Psi}_{i}=\left(\boldsymbol{\psi}_{1}^{i}, \boldsymbol{\psi}_{2}^{i}, \cdots, \boldsymbol{\psi}_{N_G}^{i}\right)^{T} \in \mathbb{R}^{N_G \times C_{de}}$.

Different from the traditional GCNs that use a fixed adjacency matrix (such as the k-nearest neighbor graph \cite{jiang2013graph}), we define a dynamic adjacency matrix ${A^G}$ as
\begin{equation}
\label{Eq:aij}
\begin{aligned}
A_{j k}^G=\frac{\exp \left(-\operatorname{ReLU}\left(\boldsymbol{w}^{T}||\boldsymbol{\psi}_{j}^i-\boldsymbol{\psi}_{k}^i||\right)\right)}{\sum_{k=1}^{N_G} \exp \left(-\operatorname{ReLU}\left(\boldsymbol{w}^{T}||\boldsymbol{\psi}_{j}^i-\boldsymbol{\psi}_{k}^i||\right)\right)},
\end{aligned}
\end{equation}
where $\boldsymbol{\psi}_{j}^i$ and $\boldsymbol{\psi}_{k}^i$ are the extracted DE features at the $j$th and $k$th channels. The linear rectification function (ReLU) is employed as an activation function here to ensure that the output of the linear operation (i.e., the dot product between the weight vector $\boldsymbol{w}$ and the node distance $||\boldsymbol{\psi}_{j}^i-\boldsymbol{\psi}_{k}^i||$) is non-negative, which introduces non-linearity into the model and improves its capacity to learn complex patterns. The weight vector $\boldsymbol{w}$ is learned by minimizing the GCN loss function as
\begin{equation}
\label{Eq:aij_loss}
\begin{aligned}
\mathcal{L}_{gcn}=\lambda\sum_{j, k=1}^{N_G}\left\|\boldsymbol{\psi}_{j}^i-\boldsymbol{\psi}_{k}^i\right\|^{2} A_{j k}^G+\|{A^G}\|_{F}^{2}.
\end{aligned}
\end{equation}
In Eq. \ref{Eq:aij_loss}, the first term quantifies the similarities between any two nodes ($\boldsymbol{\psi}_{j}^i$ and $\boldsymbol{\psi}_{k}^i$). \textcolor{black}{When the nodes exhibit dissimilar characteristics, it leads to a smaller adjacency value ($A^G_{jk}$).} As the brain network is known to exhibit sparse connectivity, we incorporate sparsity constraint into the graph learning process. A Frobenius norm term of the adjacency matrix ${A^G}$ is included in Eq. \ref{Eq:aij_loss}, which encourages the learned adjacency matrix to be sparse. $\lambda \geq 0$ is a regularization parameter that controls the trade-off between graph learning and sparsity degree.

\textcolor{black}{Based on the learned adjacency matrix ${A^G}$, the node representation could be characterized on the basis of the Chebyshev expansion of the graph Laplacian.} The Chebyshev graph convolution \cite{defferrard2016convolutional} is defined as a $\Phi-1$ degree polynomial, given as
\begin{equation}
\label{Eq:chev}
\begin{aligned}
G=\sum_{\varphi=0}^{\Phi-1} \theta_{\varphi} T_{\varphi}(\tilde{L})x_i,
\end{aligned}
\end{equation}
where $\theta \in \mathbb{R}^{\Phi}$ is a Chebyshev coefficient vector, and $x_i$ is the input sample data. $T_{\varphi}(\tilde{L}) \in \mathbb{R}^{N_G \times N_G}$ is the $\varphi$th order Chebyshev polynomial with the variable $\tilde{L}$ given as
\begin{equation}
    \tilde{L}=\frac{2}{\lambda_{\max }} L-I_{N_G},
\end{equation}
where $\lambda_{\max}$ is the maximum eigenvalue of the Laplacian matrix. $I_{N_G}$ is the identity matrix, and $L$ is the Laplacian matrix calculated as
\begin{equation}
    L=D-{A^G},
\end{equation}
where $D$ is the degree matrix. 
\textcolor{black}{
The Chebyshev polynomials are given as
\begin{equation}
    T_{\varphi}(\tilde{L})=2\tilde{L} T_{\varphi-1}(\tilde{L})-T_{\varphi-2}(\tilde{L}),
\end{equation}
where $T_{0}(\tilde{L})=1$ and $T_{1}(\tilde{L})=\tilde{L}$.}
In the final obtained graph $G$, the corresponding node representation could well capture information about the $\varphi$th order nodes of the graph and provide a richer and more comprehensive view of the graph.

\subsubsection{Graph contrastive learning}
\label{sec:GCL} 
\textcolor{black}{To further enhance feature representation, GCL is introduced to learn representations that are robust to certain transformations or augmentations of the data. This process ensures that similar instances are brought closer together in the representation space, while dissimilar instances are pushed apart.}

\textcolor{black}{Building upon the $G$ obtained in subsection \ref{sec:GCN}, two augmented graphs, denoted as $\hat{G}_i$ and $\hat{G}_j$, are generated as positive samples. We randomly drop $\zeta$\% of the nodes from $G$, following a uniform dropout probability distribution, similar to \cite{you2020graph}.} The node features of the augmented graphs are then flattened into one-dimensional feature vectors, denoted as $\{\hat{g}_1,...,\hat{g}_n\}$ and $\{\hat{g}'_1,...,\hat{g}'_n\}$, respectively. $n$ is the corresponding feature dimensionality. A feature extractor $F_S(\cdot)$ is applied to generate high-level feature representation, resulting as $\{\tilde{g}_1,...,\tilde{g}_{\dot{n}}\}$ and $\{\tilde{g}'_1,...,\tilde{g}'_{\dot{n}}\}$. $\dot{n}$ is the obtained feature dimensionality after the feature extraction. Then, a projector $P(\cdot)$ is used to further reduce the feature dimensionality, producing $\boldsymbol{z}_i$ and $\boldsymbol{z}_j$ for each augmented graph. To ensure consistency between the feature representation of the two augmented graphs generated from the same input, a contrastive learning loss $\mathcal{L}_{gcl}$ is defined as a normalized temperature-scaled cross-entropy loss, given as


\begin{equation}
\label{Eq:gcl}
\begin{aligned}
\mathcal{L}_{gcl}=-\log \frac{\exp \left(\operatorname{Sim}\left(\boldsymbol{z}_{i}, \boldsymbol{z}_{j}\right) / \tau\right)}{\sum_{k=1, k \neq i}^{B} \exp \left(\operatorname{Sim}\left(\boldsymbol{z}_{i}, \boldsymbol{z}_{k}\right) / \tau\right)},
\end{aligned}
\end{equation}
where $\operatorname{Sim}$ refers to cosine similarity and $\tau$ is a temperature parameter to adjust the feature learning performance. $\mathcal{L}_{gcl}$ encourages the similarity between $\boldsymbol{z}_{i}$ and $\boldsymbol{z}_{j}$ (positive samples) to be maximized, while pushing away the similarity between $\boldsymbol{z}_{i}$ and $\boldsymbol{z}_{k}$ (negative samples). $B$ is the batch size.

\subsection{Self-Attentive Fusion}
\label{sec:selfAtt}
\textcolor{black}{Self-attentive fusion is introduced to effectively highlight important features and assign higher weights to the source data that is in closer proximity to the target data, resulting in more informative feature representation. Furthermore, to ensure that the extracted features are discriminant for emotion recognition, a supervised classification part is incorporated into the model learning process as well.}

\subsubsection{Informative feature fusion}
The extracted non-structural features by $F_{NS}(\cdot)$ and structural features by $F_S(\cdot)$ are concatenated into a new feature representation, denoted as $\{\tilde{f}_{1}, \ldots, \tilde{f}_{\dot{m}},\tilde{g}_{1}, \ldots, \tilde{g}_{\dot{n}}\}$. \textcolor{black}{The multi-head self-attention mechanism is incorporated to emphasize the most crucial features relevant to emotions while reducing the influence of irrelevant information. It adapts to various viewpoints or focal points, autonomously determining the focus of feature extraction based on input data and task-specific requirements.} As shown in Fig. \ref{fig:standard_pipeline}, we generate three matrices $Q$, $K$, and $V$ from the input using linear transformations. Inspired from \cite{vaswani2017attention}, the attention weights are given as
\begin{equation}
\label{Eq:ATT}
\begin{aligned}
ATT(Q, K, V)=\operatorname{Softmax}\left(\frac{Q  K^{T}}{\sqrt{d}}\right) V
\end{aligned}
\end{equation}
Then, we further extend the attention mechanism to $H$ heads over the three matrices. Each matrix is divided into $H$ subspaces, termed as $Q=\{Q^{1}, \cdots, Q^{H}\}$, $K=\{K^{1}, \cdots, K^{H}\}$, $V=\{V^{1}, \cdots, V^{H}\}$. In each subspace $h \in H$, we calculate $A^{h}$ using the attention formula, given as
\begin{equation}
\label{Eq:ATT_H}
\begin{aligned}
A^{h}=A T T\left(Q^{h}, K^{h}, V^{h}\right)
\end{aligned}
\end{equation}
Finally, all $H$ representations are concatenated together to obtain the final output $MHA(X)=MHA(Q, K, V)$ for classification as 
\begin{equation}
\label{Eq:MHA}
\begin{aligned}
MHA(Q, K, V)=\operatorname{Concat}\left(A^{1}, \cdots, A^{H}\right).
\end{aligned}
\end{equation}

\subsubsection{Informative sample selection}
\label{sec:weights}
\textcolor{black}{During the model training process, we place additional attention on weighing the contribution of each labeled source data. We assign higher weights to those labeled source data that offer more valuable information for effective emotion recognition. This process helps prioritize and focus on the most informative data during the optimization process. Specifically, based on the feature representation $MHA(\cdot)$, a fully connected layer $\phi(\cdot)$ is designed as}
\begin{equation}
\label{Eq:fc1}
\begin{aligned}
\left\{\begin{array}{l}
R(X_s)=\phi(MHA(X_s))=\{r_s^1, \cdots,r_s^B\}\\
R(X_t)=\phi(MHA(X_t))=\{r_t^1, \cdots,r_t^B\},
\end{array}\right.
\end{aligned}
\end{equation}
\textcolor{black}{where $B$ represents the batch size. For each labeled source data $r_s^b$ ($1 \leq b \leq B$), we calculate the corresponding cosine similarity with all unknown target data as}
\begin{equation}
\label{Eq:sim}
\begin{aligned}
\operatorname{Sim}(r_s^b)=\frac{1}{B}\sum_{k=1}^{B}\frac{r_s^b \cdot r_t^k}{\bigl\| r_s^b \bigr\|_{2}\bigl\| r_t^k \bigr\|_2}.
\end{aligned}
\end{equation}
$\operatorname{Softmax}$ is then applied on $\{\operatorname{Sim}(r_s^1),\ldots,\operatorname{Sim}(r_s^B)\}$ for normalization. Finally, the normalized similarity weight $\operatorname{Sim}(r_s^b)$ is used to adjust the sample contribution in the multi-class cross-entropy loss function as
\begin{equation}
\label{Eq:class}
\begin{aligned}
\mathcal{L}_{c e}=-\frac{1}{B}\sum_{b=1}^{B} \sum_{c=1}^{C} y_{b}^{c} \log \left(\operatorname{Sim}(r_s^b) \cdot \hat{y}_{b}^{c}\right),
\end{aligned}
\end{equation}
\textcolor{black}{where $y_{b}^{c}$ and $\hat{y}_{b}^{c}$ denote the actual emotion label and the predicted emotion label of $b$-th labeled source data, respectively. $C$ is the total number of emotion categories.}

\color{black}
\section{Experimental Results}
\label{sec:experiment}
\subsection{Benchmark databases}
\textcolor{black}{\textcolor{black}{To assess the efficacy of the proposed DS-AGC model, comprehensive experiments are carried out on four benchmark EEG databases: SEED\cite{zheng2015investigating}, SEED-IV\cite{zheng2018emotionmeter}, SEED-V\cite{liu2021comparing} and FACED\cite{chen2023large}}. In the SEED database, three emotional states (negative, neutral, and positive) were involved, and 15 subjects (7 males and 8 females) were recruited. In the SEED-IV database, four emotional states (happiness, sadness, fear, and neutral) were selected, and 15 subjects (7 males and 8 females) were recruited. 
\textcolor{black}{In the SEED-V database, it included five emotional states (happiness, sadness, fear, disgust, and neutral), with EEG recordings from 16 subjects (6 males and 10 females).}
\textcolor{black}{In the FACED database, it included nine emotional states (amusement, inspiration,
joy, tenderness, anger, fear, disgust, sadness, and neutral), with EEG recordings from 123 subjects.}
To maintain consistency with previous research on the three benchmark databases and ensure fair comparisons, our study also utilizes the pre-computed DE features \cite{zheng2015investigating,shi2010off}.}

\subsection{Implementation details and model setting}
In the implementation, the feature extractors $F_{NS}(\cdot)$ and $F_S(\cdot)$, as well as the domain discriminator $d(\cdot)$, are composed of fully connected layers with the ReLU activation function.
\textcolor{black}{Given the input DE features with a small feature dimensionality, a lightweight network architecture would be more appropriate as discussed in prior works \cite{zhou2023pr,liang2021eegfusenet}.} Specifically, $F_{NS}(\cdot)$ is designed with 310 neurons (input layer)-64 neurons (hidden layer 1)-ReLU activation-64 neurons (hidden layer 2)-ReLU activation-64 neurons (output feature layer). The probability of node dropout \textcolor{black}{$\zeta$\%} is set to 0.2, resulting in 49 remaining channels. $F_S(\cdot)$ is designed with 245 neurons (input layer)-64 neurons (hidden layer 1)-ReLU activation-64 neurons (hidden layer 2)-ReLU activation-64 neurons (output feature layer). The domain discriminator $d(\cdot)$ is designed with 64 neurons (input layer)-64 neurons (hidden layer 1)-ReLU activation-dropout layer-64 neurons (hidden layer 2)-2 neurons (output layer) / 3 neurons (output layer)-$\operatorname{Softmax}$ activation. Gradient descent and parameter optimization are carried out using the RMSprop optimizer, with a learning rate set to 1e-3 and a batch size of 48. In the GCN architecture, we set $\varphi$ in Eq. \ref{Eq:chev} as 3 to ensure the extraction of stable graph network features. This choice strikes a balance by incorporating relevant channel features while minimizing the introduction of noise. To further fine-tune the model, we set the balancing parameter $\lambda$ in Eq. \ref{Eq:aij_loss} to 0.01. For the multi-head self-attention mechanism, we utilize $H=64$ in Eq. \ref{Eq:MHA}. This configuration allows the model to capture multiple perspectives and enhance its ability to input data. All the models are trained using the PyTorch API on an NVIDIA GeForce RTX 1080 GPU, with CUDA version 11.7. During the model training process, we solely utilize the raw target data without any label information. This approach aligns with previous EEG-based emotion recognition methods that employ a transfer learning framework, as demonstrated in studies such as \cite{li2019multisource, zhong2020eeg, wu2020transfer, li2021can}.

\subsection{Experimental protocol with incomplete labels}

\begin{figure}
\begin{center}
\includegraphics[width=0.51\textwidth]{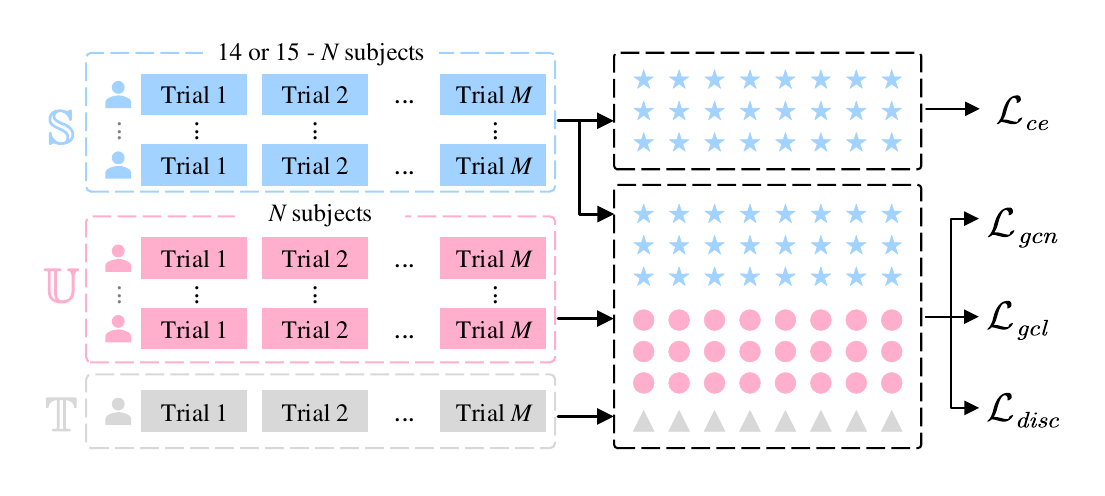}
\end{center}
\caption{\textcolor{black}{The cross-subject leave-one-subject-out cross-validation experimental protocol with incomplete labels. $\mathbb{S}$, $\mathbb{U}$, and $\mathbb{T}$ represent the labeled source domain, unlabeled source domain, and unknown target domain. For the SEED and SEED-IV databases, which contain a total of 15 subjects, $14 - N$ subjects are allocated to $\mathbb{S}$, and $N$ subjects to $\mathbb{U}$. Similarly, for the SEED-V database with 16 subjects, $15 - N$ subjects are assigned to $\mathbb{S}$, and $N$ subjects to $\mathbb{U}$. $M$ denotes the total number of trials for each subject. $\mathcal{L}_{ce}$, $\mathcal{L}_{gcn}$, $\mathcal{L}_{gcl}$, and $\mathcal{L}_{disc}$ are the classification loss, GCN loss, GCL loss, and discriminator loss, given in Eq. \ref{Eq:class}, Eq. \ref{Eq:aij_loss}, Eq. \ref{Eq:gcl}, and Eq. \ref{Eq:Tripledomainloss}. In the implementation, $\mathcal{L}_{ce}$ is calculated using only $\mathbb{S}$ (represented by blue stars), as the label information for $\mathbb{U}$ and $\mathbb{T}$ is unknown. $\mathcal{L}_{gcn}$, $\mathcal{L}_{gcl}$, and $\mathcal{L}_{disc}$, which do not depend on label information, are calculated using data from $\mathbb{S}$ (blue stars), $\mathbb{U}$ (pink circles), and $\mathbb{T}$ (gray triangles).}}
\label{fig:experimentalProtocol}
\end{figure}

\textcolor{black}{As illustrated in Fig. \ref{fig:experimentalProtocol}, we implement a strict leave-one-subject-out cross-validation experimental protocol. For the SEED and SEED-IV databases (each containing 15 subjects), we sequentially select 14 subjects as the source domain and the remaining one as the target domain for model testing. Following the same strategy, in the SEED-V database (containing 16 subjects), 15 subjects are used as the source domain, and the remaining subject is used as the target domain. \textcolor{black}{In order to} ensure each subject is rotated as the target domain, undergoing a total of 15 rounds (SEED and SEED-IV) or 16 rounds (SEED-V) of model training and testing. The final reported classification results are the average accuracy and standard deviation across 15 (SEED and SEED-IV) or 16 (SEED-V) rounds. Within the source domain, a subset of subjects ($N$ subjects) is classified as the unlabeled source domain $\mathbb{U}$, and the remaining $14-N$ subjects (SEED and SEED-IV) or $15-N$ (SEED-V) are classified as the labeled source domain $\mathbb{S}$. For example, if the first subject is assigned to $\mathbb{T}$, then subjects 2 to $2+N-1$ are assigned to $\mathbb{U}$, and subjects $2+N$ to 15 (SEED and SEED-IV) or $2+N$ to 16 (SEED-V) belong to $\mathbb{S}$. To fully evaluate the model's stability in various situations of label scarcity, we vary the $N$ value from 1 to 13 for SEED and SEED-IV, and from 1 to 14 for SEED-V in the implementation.}

During the training process of the model, the unlabeled source domain $\mathbb{U}$ is excluded from the beginning and only the labeled source domain $\mathbb{S}$ and the unknown target domain $\mathbb{T}$ participate in the training for the first $E_t$ iterations. The output layer of the domain discriminator $d(\cdot)$ has two neurons at this point, performing binary classification between the $\mathbb{S}$ and $\mathbb{T}$ domains. After the model reaches a certain stability, the unlabeled source domain $\mathbb{U}$ is added to the training, and the output layer of the domain discriminator $d(\cdot)$ has three neurons, performing ternary classification among the $\mathbb{S}$, $\mathbb{U}$, and $\mathbb{T}$ domains. Notably, only the data from the labeled source domain $\mathbb{S}$ is used for calculating the cross-entropy loss $\mathcal{L}_{c e}$ in the classifier throughout the entire process.

\subsection{Emotion recognition performance with incomplete labels}
\textcolor{black}{As shown in Table \ref{tab:seed_compareFull}, the proposed DS-AGC model consistently outperforms existing machine learning and deep learning models on the SEED database, with an average improvement of 5.83\%. Furthermore, even with an increasing number of incomplete labels, the model maintains relatively stable performance. Under the most challenging condition of extreme label scarcity ($N=13$), the DS-AGC model achieves an improvement of 7.56\%, demonstrating robust stability in situations with minimal labeled data. Table \ref{tab:seediv_compareFull} shows the experimental comparison results on the SEED-IV database. Similarly, we vary the $N$ value from 1 to 13. The corresponding average accuracy of DS-AGC on SEED-IV is $61.32\pm10.41$, with an average improvement of 0.19\%. It shows DS-AGC exhibits particularly strong adaptability and efficiency with larger $N$ value (less labeled source data). The evaluation performance on the SEED-V database is presented in Table \ref{tab:seedv_compareFull}. The DS-AGC model achieves an average accuracy of $53.87\pm11.14$, marking an average improvement of 0.48\% over existing methods. Similar to the model's performance on SEED-IV, the superiority of the DS-AGC is particularly evident in situations of acute label scarcity (when the $N$ value is larger than 7).
\textcolor{black}{These experimental results demonstrate that the proposed DS-AGC outperforms existing methods on average across different $N$ values and exhibits particular superiority when labeled source data is limited.}}

\begin{table*}[]
\begin{center}
\color{black}
\caption{The emotion recognition results on SEED database with incomplete label conditions, in terms of the average accuracy (\%) and standard deviation (\%). The model results reproduced by us are indicated by `*'.}
\label{tab:seed_compareFull}
\begin{tabular*}{\hsize}{@{}@{\extracolsep{\fill}}lccccccc@{}}
\toprule
Methods   & $N=1$ & $N=2$ & $N=3$ & $N=4$ & $N=5$ & $N=6$ & $N=7$ \\ 
\midrule
SVM*\cite{SVM1999}         &68.84$\pm$07.42	&68.64$\pm$09.92	&69.01$\pm$09.48	&69.49$\pm$09.84	&67.81$\pm$09.82	&68.00$\pm$09.87	&66.44$\pm$09.56\\
TCA*\cite{TCA2010}         &59.32$\pm$07.97	&59.51$\pm$08.06	&60.03$\pm$07.93	&59.39$\pm$08.00	&60.04$\pm$09.02	&60.09$\pm$09.30	&58.95$\pm$09.83\\
SA*\cite{SA2013}           &57.37$\pm$11.27	&61.79$\pm$08.78	&56.98$\pm$08.44	&60.58$\pm$08.29	&60.19$\pm$11.21	&56.91$\pm$09.68	&52.31$\pm$10.85\\
KPCA*\cite{KPCA1999}       &59.24$\pm$06.93	&58.59$\pm$09.10	&59.40$\pm$08.00	&59.58$\pm$08.98	&58.79$\pm$08.38	&58.75$\pm$08.58	&57.74$\pm$09.34\\
RF*\cite{breiman2001random}    &70.40$\pm$06.92	&70.99$\pm$07.51	&68.62$\pm$07.48	&70.68$\pm$06.54	&68.35$\pm$08.44	&69.18$\pm$09.81	&68.79$\pm$07.86\\
Adaboost*\cite{2006Boost}  &71.30$\pm$07.45	&72.04$\pm$08.94	&71.47$\pm$06.67	&72.49$\pm$06.98	&70.22$\pm$09.30	&69.49$\pm$08.19	&69.43$\pm$07.15\\
CORAL*\cite{CORAL2016}     &70.29$\pm$09.88	&69.42$\pm$11.63	&72.88$\pm$09.87	&70.21$\pm$11.06	&69.05$\pm$10.94	&68.75$\pm$11.01	&66.72$\pm$10.80\\
GFK*\cite{GFK2012}         &56.73$\pm$11.51	&56.86$\pm$11.80	&57.28$\pm$11.82	&57.63$\pm$12.27	&57.06$\pm$10.79	&58.04$\pm$10.44	&57.71$\pm$10.04\\
KNN*\cite{coomans1982alternative}         &54.93$\pm$11.35	&55.38$\pm$11.56	&55.23$\pm$10.36	&55.71$\pm$12.61	&56.05$\pm$09.65	&56.73$\pm$10.87	&56.74$\pm$10.22\\
DAN*\cite{Dcoral2016}      &81.05$\pm$06.81	&79.90$\pm$05.67	&82.28$\pm$07.19	&79.87$\pm$09.56	&78.13$\pm$12.09	&78.28$\pm$09.40	&76.61$\pm$10.90\\
DANN*\cite{DDC2014}        &77.11$\pm$07.63	&79.38$\pm$07.88	&79.50$\pm$09.15	&81.85$\pm$07.30	&77.17$\pm$10.11	&77.81$\pm$09.40	&76.65$\pm$08.18\\
DCORAL*\cite{Dcoral2016}   &75.13$\pm$08.33	&77.31$\pm$07.37	&78.49$\pm$07.42	&76.22$\pm$09.27	&72.39$\pm$10.63	&74.93$\pm$09.39	&73.09$\pm$08.27\\
DDC*\cite{DDC2014}         &71.67$\pm$08.01	&73.43$\pm$11.73	&77.67$\pm$05.70	&73.80$\pm$08.05	&73.39$\pm$12.92	&73.22$\pm$08.57	&73.80$\pm$08.65\\
PARSE*\cite{zhang2022parse} &79.35$\pm$06.20	&79.65$\pm$08.34	&80.45$\pm$07.78	&80.10$\pm$07.21	&79.96$\pm$07.31	&77.80$\pm$09.28	&77.31$\pm$08.86\\
MixMatch*\cite{berthelot2019mixmatch} &67.58$\pm$09.52	&71.25$\pm$06.54	&68.65$\pm$08.57	&67.50$\pm$10.04	&69.45$\pm$09.47	&69.01$\pm$09.69	&67.34$\pm$10.03\\
AdaMatch*\cite{berthelot2021adamatch} &78.14$\pm$06.18	&77.56$\pm$06.90	&78.64$\pm$06.19	&78.48$\pm$07.03	&75.53$\pm$05.96	&76.12$\pm$06.16	&73.95$\pm$06.53\\
FlexMatch*\cite{zhang2021flexmatch} &77.89$\pm$07.24	&78.09$\pm$07.17	&79.01$\pm$06.17	&78.92$\pm$06.18	&78.84$\pm$07.49	&78.47$\pm$07.54	&77.32$\pm$07.28\\
SoftMatch*\cite{chen2023softmatch} &79.09$\pm$06.25	&78.87$\pm$06.49	&79.77$\pm$05.92	&80.00$\pm$05.72	&78.73$\pm$06.71	&78.65$\pm$07.11	&78.21$\pm$06.51\\
\midrule
\multirow{2}{*}{\textbf{DS-AGC}}	&\textbf{85.98$\pm$06.21}	&\textbf{87.37$\pm$06.19}	&\textbf{86.38$\pm$07.25}	&\textbf{85.27$\pm$06.32}	&\textbf{82.70$\pm$06.37}	&\textbf{83.24$\pm$08.29}	&\textbf{84.08$\pm$06.49}\\ 
\hspace*{\fill}& \textcolor{red}{\scriptsize \textbf{(+4.93)}} & \textcolor{red}{\scriptsize \textbf{(+7.47)}} & \textcolor{red}{\scriptsize \textbf{(+4.10)}} & \textcolor{red}{\scriptsize \textbf{(+3.42)}} & \textcolor{red}{\scriptsize \textbf{(+2.74)}} & \textcolor{red}{\scriptsize \textbf{(+4.59)}} & \textcolor{red}{\scriptsize \textbf{(+5.87)}}\\
\midrule
\end{tabular*}
\vspace{0mm}
\begin{tabular*}{\hsize}{@{}@{\extracolsep{\fill}}lccccccc@{}}
\midrule
Methods   & $N=8$ & $N=9$ & $N=10$ & $N=11$ & $N=12$  & $N=13$  & Average \\ 
\midrule
SVM*\cite{SVM1999}         &63.27$\pm$11.33	&62.17$\pm$10.51	&60.49$\pm$11.54	&62.10$\pm$12.15	&63.57$\pm$11.42	&57.63$\pm$12.36	&65.19$\pm$11.15\\
TCA*\cite{TCA2010}         &59.01$\pm$08.89	&60.30$\pm$07.75	&60.37$\pm$05.11	&61.35$\pm$06.07	&62.43$\pm$08.94	&62.31$\pm$13.00	&60.24$\pm$08.73\\
SA*\cite{SA2013}           &55.05$\pm$07.85	&58.42$\pm$09.22	&56.02$\pm$08.78	&54.31$\pm$10.17	&53.66$\pm$12.64	&53.34$\pm$10.14	&56.69$\pm$10.32\\
KPCA*\cite{KPCA1999}       &57.22$\pm$10.10	&55.13$\pm$10.31	&53.54$\pm$09.84	&53.98$\pm$10.78	&54.54$\pm$11.68	&50.43$\pm$11.97	&56.71$\pm$10.06\\
RF*\cite{breiman2001random}    &66.03$\pm$08.00	&65.70$\pm$07.78	&63.48$\pm$07.93	&61.73$\pm$09.17	&59.73$\pm$10.63	&57.05$\pm$11.95	&66.21$\pm$09.63\\
Adaboost*\cite{2006Boost}  &66.99$\pm$07.65	&67.02$\pm$07.26	&66.43$\pm$08.35	&65.22$\pm$11.56	&54.17$\pm$15.92	&34.47$\pm$00.00	&65.44$\pm$13.37\\
CORAL*\cite{CORAL2016}     &64.89$\pm$11.40	&63.69$\pm$10.58	&60.45$\pm$09.89	&64.49$\pm$12.64	&61.74$\pm$10.88	&57.52$\pm$11.79	&66.16$\pm$11.83\\
GFK*\cite{GFK2012}         &57.35$\pm$08.96	&56.75$\pm$09.68	&54.11$\pm$09.65	&55.70$\pm$11.34	&56.86$\pm$09.77	&54.97$\pm$14.08	&56.70$\pm$11.10\\
KNN*\cite{coomans1982alternative}         &56.42$\pm$09.39	&55.35$\pm$09.91	&52.45$\pm$10.88	&54.19$\pm$11.33	&56.07$\pm$10.42	&53.62$\pm$13.78	&55.30$\pm$11.11\\
DAN*\cite{Dcoral2016}      &74.00$\pm$09.51	&73.35$\pm$12.48	&72.80$\pm$11.01	&73.79$\pm$12.11	&72.20$\pm$10.18	&70.37$\pm$18.13	&76.35$\pm$11.43\\
DANN*\cite{DDC2014}        &73.74$\pm$09.19	&70.38$\pm$10.46	&70.31$\pm$09.26	&71.15$\pm$13.76	&67.87$\pm$15.49	&66.65$\pm$19.71	&74.58$\pm$12.08\\
DCORAL*\cite{Dcoral2016}   &71.30$\pm$07.11	&69.10$\pm$08.74	&70.27$\pm$11.30	&70.76$\pm$12.95	&68.54$\pm$10.81	&62.45$\pm$20.62	&72.31$\pm$11.51\\
DDC*\cite{DDC2014}         &69.01$\pm$09.44	&68.75$\pm$08.45	&70.28$\pm$10.82	&71.50$\pm$08.19	&65.43$\pm$09.21	&64.98$\pm$15.04	&71.30$\pm$10.46\\
PARSE*\cite{zhang2022parse} &75.30$\pm$09.64	&75.94$\pm$08.88	&70.86$\pm$08.88	&72.09$\pm$08.76	&66.69$\pm$07.00	&65.27$\pm$10.51	&75.44$\pm$09.79\\
MixMatch*\cite{berthelot2019mixmatch} &64.85$\pm$11.89	&63.91$\pm$11.85	&60.13$\pm$08.67	&59.49$\pm$08.96	&55.16$\pm$08.27	&51.23$\pm$08.62	&64.27$\pm$11.14\\
AdaMatch*\cite{berthelot2021adamatch} &70.18$\pm$09.69	&68.17$\pm$09.49	&65.74$\pm$10.43	&66.24$\pm$11.38	&64.35$\pm$10.89	&58.14$\pm$09.35	&71.63$\pm$10.54\\
FlexMatch*\cite{zhang2021flexmatch} &74.63$\pm$07.99	&72.92$\pm$07.27	&71.29$\pm$08.04	&70.14$\pm$09.15	&70.14$\pm$11.27	&64.94$\pm$12.52	&74.82$\pm$09.38\\
SoftMatch*\cite{chen2023softmatch} &74.97$\pm$07.22	&72.84$\pm$07.13	&72.10$\pm$07.39	&69.64$\pm$09.20	&69.44$\pm$11.57	&64.18$\pm$10.87	&75.12$\pm$09.15\\
\midrule
\multirow{2}{*}{\textbf{DS-AGC}}	&\textbf{80.55$\pm$06.84}	&\textbf{80.62$\pm$06.05}	&\textbf{78.35$\pm$07.48}	&\textbf{77.20$\pm$07.40}	&\textbf{78.72$\pm$09.58}	&\textbf{77.93$\pm$12.87}	&\textbf{82.18$\pm$08.41}\\ 
\hspace*{\fill}& \textcolor{red}{\scriptsize \textbf{(+5.25)}} & \textcolor{red}{\scriptsize \textbf{(+4.68)}} & \textcolor{red}{\scriptsize \textbf{(+5.55)}} & \textcolor{red}{\scriptsize \textbf{(+3.41)}} & \textcolor{red}{\scriptsize \textbf{(+6.52)}} & \textcolor{red}{\scriptsize \textbf{(+7.56)}} & \textcolor{red}{\scriptsize \textbf{(+5.83)}}\\
\bottomrule
\end{tabular*}
\end{center}
\end{table*}

\begin{table*}[]
\begin{center}
\color{black}
\caption{The emotion recognition results on SEED-IV database with incomplete label conditions, in terms of the average accuracy (\%) and standard deviation (\%). The model results reproduced by us are indicated by `*'.}
\label{tab:seediv_compareFull}
\begin{tabular*}{\hsize}{@{}@{\extracolsep{\fill}}lccccccc@{}}
\toprule
Methods   & $N=1$ & $N=2$ & $N=3$ & $N=4$ & $N=5$ & $N=6$ & $N=7$ \\ 
\midrule
SVM*\cite{SVM1999}	         &47.47$\pm$13.59	&47.52$\pm$13.48	&45.35$\pm$13.16	&43.28$\pm$10.00	&45.51$\pm$12.74	&44.69$\pm$14.10	&48.61$\pm$15.16\\
TCA*\cite{TCA2010}	         &43.58$\pm$10.74	&44.15$\pm$10.73	&43.19$\pm$08.76	&42.24$\pm$11.14	&43.71$\pm$10.85	&40.69$\pm$08.88	&41.51$\pm$09.06\\
SA*\cite{SA2013}	         &40.82$\pm$14.33	&38.87$\pm$09.93	&40.15$\pm$10.61	&40.34$\pm$13.42	&33.07$\pm$08.48	&35.24$\pm$11.01	&36.19$\pm$14.92\\
KPCA*\cite{KPCA1999}	     &41.41$\pm$08.02	&41.05$\pm$07.72	&41.18$\pm$08.85	&38.53$\pm$10.62	&39.20$\pm$08.80	&38.31$\pm$10.38	&41.25$\pm$11.21\\
RF*\cite{breiman2001random}	     &53.65$\pm$14.12	&51.64$\pm$14.80	&49.97$\pm$14.29	&45.94$\pm$13.38	&50.28$\pm$13.56	&46.33$\pm$14.64	&47.88$\pm$15.83\\
Adaboost*\cite{2006Boost}	 &54.97$\pm$15.55	&51.82$\pm$11.41	&52.21$\pm$11.82	&50.96$\pm$11.89	&54.40$\pm$15.43	&51.76$\pm$14.46	&53.05$\pm$13.27\\
CORAL*\cite{CORAL2016}	     &50.88$\pm$15.00	&46.39$\pm$13.46	&45.17$\pm$13.29	&46.74$\pm$13.13	&47.86$\pm$14.55	&47.11$\pm$15.74	&47.75$\pm$15.33\\
GFK*\cite{GFK2012}	         &41.62$\pm$09.07	&42.54$\pm$06.90	&42.24$\pm$06.44	&40.93$\pm$06.14	&40.73$\pm$07.71	&40.79$\pm$07.87	&42.55$\pm$09.49\\
KNN*\cite{coomans1982alternative}	         &37.67$\pm$09.33	&40.04$\pm$05.55	&39.50$\pm$08.22	&39.57$\pm$06.00	&37.01$\pm$07.66	&36.72$\pm$06.06	&38.73$\pm$06.73\\
DAN*\cite{Dcoral2016}	     &57.83$\pm$07.73	&59.35$\pm$08.58	&57.47$\pm$09.01	&53.59$\pm$09.25	&55.22$\pm$09.82	&54.15$\pm$08.26	&51.94$\pm$09.97\\
DANN*\cite{DDC2014}	         &57.65$\pm$09.73	&56.89$\pm$07.99	&56.81$\pm$09.11	&54.95$\pm$10.85	&56.14$\pm$10.54	&55.87$\pm$09.76	&56.72$\pm$11.18\\
DCORAL*\cite{Dcoral2016}	 &58.71$\pm$08.36	&56.81$\pm$07.80	&58.43$\pm$08.54	&54.57$\pm$09.35	&55.18$\pm$09.81	&54.60$\pm$10.53	&53.19$\pm$11.17\\
DDC*\cite{DDC2014}	         &55.90$\pm$08.07	&54.77$\pm$08.56	&53.40$\pm$09.45	&51.73$\pm$08.39	&50.61$\pm$10.33	&51.35$\pm$09.02	&51.02$\pm$10.92\\
PARSE*\cite{zhang2022parse} &66.57$\pm$10.81	&66.19$\pm$09.71	&66.38$\pm$10.34	&64.28$\pm$10.13	&63.68$\pm$09.40	&61.12$\pm$10.74	&60.87$\pm$12.27\\
MixMatch*\cite{berthelot2019mixmatch} &61.98$\pm$08.85	&62.58$\pm$09.42	&61.60$\pm$07.71	&60.72$\pm$09.11	&61.42$\pm$07.94	&59.55$\pm$08.63	&57.85$\pm$08.44\\
AdaMatch*\cite{berthelot2021adamatch} &65.22$\pm$09.32	&64.26$\pm$09.35	&65.79$\pm$09.31	&62.23$\pm$09.91	&63.42$\pm$10.25	&62.14$\pm$09.99	&61.12$\pm$10.26\\
FlexMatch*\cite{zhang2021flexmatch} &62.77$\pm$08.93	&62.75$\pm$08.83	&62.68$\pm$09.06	&61.13$\pm$09.15	&61.06$\pm$09.41	&60.29$\pm$08.75	&59.85$\pm$09.37\\
SoftMatch*\cite{chen2023softmatch} &62.46$\pm$09.14	&62.57$\pm$08.87	&62.51$\pm$09.15	&61.14$\pm$09.41	&60.74$\pm$09.71	&60.94$\pm$08.74	&59.93$\pm$09.50\\
\midrule
\multirow{2}{*}{\textbf{DS-AGC}}	&\textbf{65.79$\pm$09.55}	&\textbf{66.00$\pm$07.93}	&\textbf{65.30$\pm$06.99}	&\textbf{63.15$\pm$08.30}	&\textbf{63.24$\pm$10.15}	&\textbf{62.40$\pm$09.65}	&\textbf{61.21$\pm$10.35}\\ 
\hspace*{\fill}& \textcolor{red}{\scriptsize \textbf{(-0.78)}} & \textcolor{red}{\scriptsize \textbf{(-0.19)}} & \textcolor{red}{\scriptsize \textbf{(-1.08)}} & \textcolor{red}{\scriptsize \textbf{(-1.13)}} & \textcolor{red}{\scriptsize \textbf{(-0.44)}} & \textcolor{red}{\scriptsize \textbf{(+0.26)}} & \textcolor{red}{\scriptsize \textbf{(+0.09)}}\\
\midrule
\end{tabular*}
\vspace{0mm}
\begin{tabular*}{\hsize}{@{}@{\extracolsep{\fill}}lccccccc@{}}
\midrule
Methods   & $N=8$ & $N=9$ & $N=10$ & $N=11$ & $N=12$  & $N=13$  & Average \\ 
\midrule
SVM*\cite{SVM1999}	         &47.19$\pm$15.04	&46.93$\pm$14.06	&46.60$\pm$11.47	&46.18$\pm$11.23	&44.36$\pm$16.42	&42.07$\pm$11.60	&45.83$\pm$13.50 \\
TCA*\cite{TCA2010}	         &42.45$\pm$09.05	&41.31$\pm$09.23	&42.64$\pm$10.68	&37.97$\pm$10.84	&39.30$\pm$12.12	&40.06$\pm$10.75	&41.75$\pm$10.45\\
SA*\cite{SA2013}	         &34.28$\pm$13.13	&39.69$\pm$13.06	&34.85$\pm$14.87	&43.07$\pm$13.75	&38.52$\pm$12.78	&32.82$\pm$08.54	&37.53$\pm$12.84\\
KPCA*\cite{KPCA1999}	     &38.22$\pm$11.35	&38.48$\pm$11.19	&40.27$\pm$14.15	&34.49$\pm$10.61	&36.80$\pm$11.32	&37.12$\pm$13.88	&38.95$\pm$11.00\\
RF*\cite{breiman2001random}	     &44.52$\pm$11.25	&45.74$\pm$16.01	&47.21$\pm$14.39	&45.34$\pm$13.38	&46.63$\pm$13.10	&40.36$\pm$12.84	&47.35$\pm$14.44\\
Adaboost*\cite{2006Boost}	 &51.48$\pm$11.56	&50.79$\pm$14.56	&51.36$\pm$13.33	&50.66$\pm$15.36	&48.16$\pm$12.26	&43.33$\pm$12.09	&51.15$\pm$13.72\\
CORAL*\cite{CORAL2016}	     &48.59$\pm$13.99	&46.92$\pm$15.07	&47.66$\pm$12.11	&44.66$\pm$12.47	&44.14$\pm$14.35	&43.83$\pm$10.56	&46.75$\pm$14.01\\
GFK*\cite{GFK2012}	         &42.63$\pm$08.23	&43.07$\pm$08.46	&40.59$\pm$11.63	&37.95$\pm$10.90	&43.60$\pm$10.64	&44.08$\pm$12.55	&41.79$\pm$09.29\\
KNN*\cite{coomans1982alternative}	         &38.24$\pm$08.36	&39.47$\pm$08.97	&36.75$\pm$13.21	&36.15$\pm$12.60	&42.32$\pm$10.79	&42.18$\pm$12.26	&38.80$\pm$09.47\\
DAN*\cite{Dcoral2016}	     &54.49$\pm$08.19	&53.34$\pm$10.22	&51.66$\pm$10.48	&51.33$\pm$12.00	&51.70$\pm$10.89	&49.34$\pm$12.07	&53.95$\pm$10.21\\
DANN*\cite{DDC2014}	         &53.68$\pm$12.30	&55.28$\pm$09.85	&51.73$\pm$11.13	&51.91$\pm$10.80	&50.84$\pm$13.06	&47.85$\pm$12.62	&54.33$\pm$11.14\\
DCORAL*\cite{Dcoral2016}	 &53.85$\pm$09.01	&53.11$\pm$10.69	&52.17$\pm$09.21	&49.39$\pm$09.84	&48.48$\pm$09.85	&49.57$\pm$11.02	&53.70$\pm$10.17\\
DDC*\cite{DDC2014}	         &50.01$\pm$09.46	&52.63$\pm$10.11	&53.10$\pm$12.24	&48.76$\pm$09.45	&50.72$\pm$12.97	&46.91$\pm$10.91	&51.61$\pm$10.35\\
PARSE*\cite{zhang2022parse} &60.49$\pm$11.57	&60.61$\pm$10.32	&60.61$\pm$12.18	&55.97$\pm$09.16	&55.49$\pm$10.66	&52.50$\pm$08.37	&61.13$\pm$11.31\\
MixMatch*\cite{berthelot2019mixmatch} &59.02$\pm$08.43	&57.65$\pm$07.72	&54.94$\pm$08.76	&53.53$\pm$06.19	&53.00$\pm$10.59	&47.55$\pm$06.82	&57.80$\pm$09.44\\
AdaMatch*\cite{berthelot2021adamatch} &60.42$\pm$10.90	&59.77$\pm$11.17	&59.11$\pm$10.63	&55.61$\pm$10.72	&55.14$\pm$11.86	&52.54$\pm$08.26	&60.52$\pm$10.91\\
FlexMatch*\cite{zhang2021flexmatch} &59.25$\pm$08.53	&59.83$\pm$08.81	&57.49$\pm$11.27	&55.84$\pm$09.54	&55.27$\pm$09.95	&53.86$\pm$08.69	&59.39$\pm$09.70\\
SoftMatch*\cite{chen2023softmatch} &59.01$\pm$08.61	&59.54$\pm$08.76	&57.90$\pm$10.54	&56.19$\pm$09.51	&55.32$\pm$10.06	&53.83$\pm$08.89	&59.39$\pm$09.71\\
\midrule
\multirow{2}{*}{\textbf{DS-AGC}}	&\textbf{62.17$\pm$10.89}	&\textbf{62.54$\pm$09.60}	&\textbf{59.55$\pm$10.53}	&\textbf{58.75$\pm$11.33}	&\textbf{57.09$\pm$08.52}	&\textbf{50.00$\pm$09.31}	&\textbf{61.32$\pm$10.41}\\ 
\hspace*{\fill}& \textcolor{red}{\scriptsize \textbf{(+1.68)}} & \textcolor{red}{\scriptsize \textbf{(+1.93)}} & \textcolor{red}{\scriptsize \textbf{(-1.06)}} & \textcolor{red}{\scriptsize \textbf{(+2.56)}} & \textcolor{red}{\scriptsize \textbf{(+1.60)}} & \textcolor{red}{\scriptsize \textbf{(-3.86)}} & \textcolor{red}{\scriptsize \textbf{(+0.19)}}\\
\bottomrule
\end{tabular*}
\end{center}
\end{table*}

\begin{table*}[]
\begin{center}
\color{black}
\caption{The emotion recognition results on SEED-V database with incomplete label conditions, in terms of the average accuracy (\%) and standard deviation (\%). The model results reproduced by us are indicated by `*'.}
\label{tab:seedv_compareFull}
\begin{tabular*}{\hsize}{@{}@{\extracolsep{\fill}}lcccccccc@{}}
\toprule
Methods   & $N=1$ & $N=2$ & $N=3$ & $N=4$ & $N=5$ & $N=6$ & $N=7$ & $N=8$\\ 
\midrule
SVM*\cite{SVM1999}	&48.00$\pm$23.23	&47.05$\pm$22.07	&45.95$\pm$20.28	&44.39$\pm$18.99	&41.31$\pm$22.76	&43.01$\pm$19.53	&45.64$\pm$18.06	&38.74$\pm$19.03\\
TCA*\cite{TCA2010}	&37.23$\pm$14.32	&36.48$\pm$16.61	&34.11$\pm$12.47	&35.45$\pm$14.52	&34.11$\pm$12.47	&34.11$\pm$12.47	&32.53$\pm$13.06	&34.30$\pm$13.10\\
SA*\cite{SA2013}	&29.92$\pm$07.12	&27.24$\pm$10.30	&28.22$\pm$07.87	&31.07$\pm$06.56	&28.61$\pm$07.14	&33.09$\pm$10.59	&29.54$\pm$09.15	&34.32$\pm$10.96\\
KPCA*\cite{KPCA1999}&31.61$\pm$16.35	&32.38$\pm$16.32	&33.00$\pm$15.47	&29.97$\pm$17.24	&29.07$\pm$15.26	&30.20$\pm$15.06	&28.46$\pm$13.71	&29.06$\pm$12.25\\
RF*\cite{breiman2001random}	     &45.94$\pm$16.70	&44.91$\pm$17.56	&43.12$\pm$21.45	&41.32$\pm$14.33	&46.41$\pm$17.39	&40.19$\pm$16.40	&43.58$\pm$15.69	&41.05$\pm$15.76\\
Adaboost*\cite{2006Boost}	 &48.36$\pm$22.10	&47.36$\pm$21.51	&45.65$\pm$19.31	&48.74$\pm$20.92	&47.49$\pm$19.22	&47.87$\pm$14.71	&44.35$\pm$16.63	&43.74$\pm$13.35\\
CORAL*\cite{CORAL2016}	     &52.41$\pm$25.20	&46.80$\pm$21.78	&46.84$\pm$21.56	&45.58$\pm$19.43	&40.74$\pm$22.74	&42.33$\pm$20.44	&46.96$\pm$20.19	&43.02$\pm$23.13\\
GFK*\cite{GFK2012}	&36.12$\pm$15.53	&33.58$\pm$15.60	&34.21$\pm$16.46	&33.18$\pm$15.44	&33.27$\pm$14.30	&33.21$\pm$14.04	&32.89$\pm$12.87	&32.35$\pm$13.04\\
KNN*\cite{coomans1982alternative}	 &33.81$\pm$14.43	&31.36$\pm$14.94	&31.80$\pm$15.91	&32.23$\pm$14.78	&32.73$\pm$13.50	&32.30$\pm$13.75	&31.87$\pm$12.95	&33.09$\pm$11.96\\
DAN*\cite{Dcoral2016}	    &50.39$\pm$11.71	&51.58$\pm$14.31	&49.57$\pm$11.82	&51.22$\pm$14.17	&49.39$\pm$13.14	&47.53$\pm$14.27	&49.61$\pm$14.26	&48.20$\pm$15.28\\
DANN*\cite{DDC2014}	        &53.00$\pm$15.31	&50.96$\pm$14.36	&48.67$\pm$14.12	&50.58$\pm$15.94	&50.09$\pm$14.41	&50.40$\pm$17.87	&47.79$\pm$15.11	&44.63$\pm$14.98\\
DCORAL*\cite{Dcoral2016}	&50.67$\pm$12.41	&50.25$\pm$14.12	&49.34$\pm$12.37	&47.82$\pm$13.56	&46.82$\pm$13.05	&48.64$\pm$15.16	&46.97$\pm$14.02	&47.30$\pm$15.56\\
DDC*\cite{DDC2014}	        &50.47$\pm$10.99	&48.31$\pm$13.62	&47.21$\pm$11.38	&47.08$\pm$12.23	&48.33$\pm$12.72	&44.02$\pm$12.96	&45.68$\pm$12.65	&45.80$\pm$13.45\\
PARSE*\cite{zhang2022parse} &59.24$\pm$12.12	&60.05$\pm$12.87	&59.60$\pm$12.66	&58.15$\pm$12.29	&56.39$\pm$12.63	&55.35$\pm$13.55	&56.38$\pm$14.84	&53.65$\pm$16.22\\
MixMatch*\cite{berthelot2019mixmatch} &57.22$\pm$10.77	&57.20$\pm$11.91	&56.39$\pm$11.18	&54.63$\pm$11.22	&54.46$\pm$11.55	&53.30$\pm$12.63	&52.86$\pm$11.28	&51.12$\pm$12.39\\
AdaMatch*\cite{berthelot2021adamatch} &60.02$\pm$12.47	&59.02$\pm$12.32	&57.73$\pm$12.62	&57.13$\pm$12.32	&55.43$\pm$12.72	&55.46$\pm$13.94	&54.28$\pm$14.00	&51.28$\pm$15.56\\
FlexMatch*\cite{zhang2021flexmatch} &56.13$\pm$13.93	&56.48$\pm$14.10	&55.12$\pm$13.93	&54.92$\pm$13.58	&54.78$\pm$14.33	&56.17$\pm$14.02	&55.53$\pm$12.74	&53.48$\pm$12.61\\
SoftMatch*\cite{chen2023softmatch} &56.63$\pm$14.06	&55.71$\pm$14.01	&54.96$\pm$14.26	&54.66$\pm$13.91	&55.29$\pm$14.06	&56.20$\pm$14.08	&55.28$\pm$13.23	&53.18$\pm$12.68\\
\midrule
\multirow{2}{*}{\textbf{DS-AGC}}	&\textbf{60.71$\pm$08.99}	&\textbf{59.40$\pm$09.99}	&\textbf{55.86$\pm$10.10}	&\textbf{55.19$\pm$09.34}	&\textbf{54.60$\pm$08.52}	&\textbf{54.01$\pm$09.29}	&\textbf{55.63$\pm$12.25}  &\textbf{54.63$\pm$08.96}\\ 
\hspace*{\fill}& \textcolor{red}{\scriptsize \textbf{(+0.69)}} & \textcolor{red}{\scriptsize \textbf{(-0.65)}} & \textcolor{red}{\scriptsize \textbf{(-3.74)}} & \textcolor{red}{\scriptsize \textbf{(-2.96)}} & \textcolor{red}{\scriptsize \textbf{(-1.79)}} & \textcolor{red}{\scriptsize \textbf{(-2.19)}} & \textcolor{red}{\scriptsize \textbf{(-0.75)}} & \textcolor{red}{\scriptsize \textbf{(+0.98)}}\\
\midrule
\end{tabular*}
\vspace{0mm}
\begin{tabular*}{\hsize}{@{}@{\extracolsep{\fill}}lccccccc@{}}
\midrule
Methods   & $N=9$ & $N=10$ & $N=11$ & $N=12$  & $N=13$ & $N=14$ & Average \\ 
\midrule
SVM*\cite{SVM1999}	&44.29$\pm$19.83	&44.74$\pm$19.73	&40.58$\pm$19.05	&36.69$\pm$20.27	&33.38$\pm$17.54	&32.13$\pm$17.09	&41.19$\pm$20.52\\
TCA*\cite{TCA2010}	&30.80$\pm$12.54	&34.05$\pm$12.82	&33.23$\pm$12.89	&29.09$\pm$16.43	&35.36$\pm$17.70	&30.93$\pm$11.29	&33.50$\pm$14.43\\
SA*\cite{SA2013}	&30.35$\pm$12.68	&33.05$\pm$13.42	&35.09$\pm$13.63	&32.91$\pm$10.64	&30.54$\pm$14.23	&25.76$\pm$12.20	&31.03$\pm$11.26\\
KPCA*\cite{KPCA1999}&28.03$\pm$11.00	&27.65$\pm$10.44	&29.05$\pm$09.53	&27.98$\pm$08.40	&26.64$\pm$11.81	&25.89$\pm$09.11	&28.86$\pm$13.00\\
RF*\cite{breiman2001random}	    &38.46$\pm$16.20	&40.34$\pm$13.89	&36.44$\pm$14.73	&38.47$\pm$19.91	&34.07$\pm$13.49	&31.62$\pm$13.06	&40.09$\pm$16.77\\
Adaboost*\cite{2006Boost}	&42.97$\pm$14.28	&43.39$\pm$15.20	&39.51$\pm$15.35	&34.67$\pm$14.19	&33.92$\pm$11.02	&34.38$\pm$13.21	&42.24$\pm$17.41\\
CORAL*\cite{CORAL2016}	    &46.58$\pm$23.77	&44.71$\pm$18.82	&42.00$\pm$20.11	&38.08$\pm$21.46	&32.95$\pm$17.01	&29.80$\pm$15.63	&42.15$\pm$21.77\\
GFK*\cite{GFK2012}	        &31.15$\pm$12.26	&32.39$\pm$11.80	&32.40$\pm$10.31	&29.69$\pm$11.70	&30.21$\pm$13.64	&31.75$\pm$09.80	&32.28$\pm$13.26\\
KNN*\cite{coomans1982alternative}	   &32.66$\pm$12.87	&33.01$\pm$11.68	&32.37$\pm$11.13	&29.69$\pm$11.89	&29.79$\pm$13.06	&30.73$\pm$10.52	&31.81$\pm$12.97\\
DAN*\cite{Dcoral2016}	    &47.98$\pm$14.59	&48.78$\pm$16.26	&44.72$\pm$14.80	&41.28$\pm$13.05	&39.05$\pm$13.14	&36.72$\pm$14.76	&47.64$\pm$14.43\\
DANN*\cite{DDC2014}	        &49.72$\pm$16.13	&47.62$\pm$15.76	&45.05$\pm$12.37	&40.61$\pm$14.66	&38.57$\pm$14.69	&38.44$\pm$13.11	&47.52$\pm$15.64\\
DCORAL*\cite{Dcoral2016}	 &46.26$\pm$14.30	&49.29$\pm$14.93	&45.81$\pm$13.47	&42.14$\pm$13.05	&41.55$\pm$12.27	&38.39$\pm$15.65	&47.14$\pm$14.01\\
DDC*\cite{DDC2014}	        &46.99$\pm$13.02	&44.62$\pm$14.47	&41.33$\pm$10.78	&37.48$\pm$12.68	&39.35$\pm$12.57	&33.30$\pm$11.80	&45.13$\pm$13.12\\
PARSE*\cite{zhang2022parse} &53.00$\pm$15.65	&52.73$\pm$15.31	&51.54$\pm$15.80 &46.95$\pm$14.22	&46.02$\pm$13.99	&38.39$\pm$13.31	&53.39$\pm$15.22\\
MixMatch*\cite{berthelot2019mixmatch} &50.70$\pm$12.90	&50.54$\pm$12.81	&48.66$\pm$10.98	&45.50$\pm$10.14	&44.97$\pm$10.33	&39.54$\pm$10.49	&51.22$\pm$12.52\\
AdaMatch*\cite{berthelot2021adamatch} &51.35$\pm$15.90	&51.48$\pm$14.55	&49.70$\pm$13.83	&46.44$\pm$13.60	&43.72$\pm$11.05	&36.52$\pm$13.38	&52.11$\pm$14.88\\
FlexMatch*\cite{zhang2021flexmatch} &51.88$\pm$13.02	&52.37$\pm$14.00	&50.37$\pm$13.22	&48.44$\pm$11.73	&47.41$\pm$12.85	&40.23$\pm$13.02	&52.38$\pm$14.08\\
SoftMatch*\cite{chen2023softmatch} &51.80$\pm$13.33	&52.08$\pm$14.51	&50.33$\pm$12.78	&48.26$\pm$11.77	&48.18$\pm$12.44	&40.83$\pm$12.95	&52.38$\pm$14.13\\
\midrule
\multirow{2}{*}{\textbf{DS-AGC}}	&\textbf{54.78$\pm$09.52}	&\textbf{54.54$\pm$13.61}	&\textbf{54.27$\pm$13.02}	&\textbf{50.74$\pm$07.42}	&\textbf{47.47$\pm$09.59}	&\textbf{42.39$\pm$10.72}	&\textbf{53.87$\pm$11.14}\\ 
\hspace*{\fill}& \textcolor{red}{\scriptsize \textbf{(+1.78)}} & \textcolor{red}{\scriptsize \textbf{(+1.81)}} & \textcolor{red}{\scriptsize \textbf{(+2.73)}} & \textcolor{red}{\scriptsize \textbf{(+2.30)}} & \textcolor{red}{\scriptsize \textbf{(-0.71)}} & \textcolor{red}{\scriptsize \textbf{(+1.56)}} & \textcolor{red}{\scriptsize \textbf{(+0.48)}}\\
\bottomrule
\end{tabular*}
\end{center}
\end{table*}

\subsection{\textcolor{black}{Further validation on the FACED dataset}} 
\textcolor{black}{To further assess the effectiveness of the proposed DS-AGC semi-supervised learning model in handling large amounts of unlabeled data, we also validate it with the FACED dataset \cite{chen2023large}. For a total of 123 subjects, each subject underwent 28 trials corresponding to 9 different emotions: amusement, inspiration, joy, tenderness (categorized as positive emotions), anger, fear, disgust, sadness (categorized as negative emotions), and a neutral state. Following standard EEG preprocessing, similar to the SEED series database, DE features were extracted.}

\textcolor{black}{Here, we evaluate the FACED database under a ten-fold cross-validation method. Specifically, the total of 123 subjects is randomly divided into ten groups. In each iteration, one group serves as the unknown target domain for testing, while the remaining nine groups serve as the source domain\textcolor{black}{, and this process is repeated for ten rounds}. This ensures that each subset is used as the test set once. In the source domain, the training data is further divided into labeled ($\mathbb{S}$) and unlabeled ($\mathbb{U}$) domains according to a specific ratio. To fully verify the model stability under different conditions of label scarcity, three semi-supervised cases are considered. (1) 66\% of the data is assigned as labeled source data and 33\% as unlabeled source data ($\mathbb{S}:\mathbb{U}=2:1$). (2) 50\% of the data is assigned as labeled source data and 50\% as unlabeled source data ($\mathbb{S}:\mathbb{U}=1:1$). (3) 33\% of the data is assigned as labeled source data and 66\% as unlabeled source data ($\mathbb{S}:\mathbb{U}=1:2$). The third case is the most challenging, as it has the smallest amount of labeled source data. Additionally, two types of emotional classification tasks are conducted: a three-category classification (positive, neutral, and negative emotions) and a nine-category classification (amusement, inspiration, joy, tenderness, anger, fear, disgust, sadness, and neutral), denoted as $Label_3$ and $Label_9$, respectively. As shown in Table \ref{tab:FACEDdata}, the experimental results indicate that, compared to the existing methods, the proposed DS-AGC model achieves superior performance, with an average improvement of 0.43\% in three-class classification results and 0.64\% in nine-class classification results.}

\begin{table*}[h]
\setlength{\tabcolsep}{0.45em}
\begin{center}
\color{black}
\caption{\textcolor{black}{Experimental results on the FACED database using cross-subject ten-fold cross-validation protocol with incomplete labels. $Label_3$: three-category classification (positive, neutral, and negative emotions); $Label_9$: nine-category classification (amusement, inspiration, joy, tenderness, anger, fear, disgust, sadness, and neutral). The model results reproduced by us are indicated by `*'.}}
\label{tab:FACEDdata}
\scalebox{1}{
\begin{tabular*}{\hsize}{@{}@{\extracolsep{\fill}}ccccccc@{}}
\toprule
Session & $\mathbb{S} : \mathbb{U}$ & MixMatch*\cite{berthelot2019mixmatch} & AdaMatch*\cite{berthelot2021adamatch} & FlexMatch*\cite{zhang2021flexmatch} & SoftMatch*\cite{chen2023softmatch} & \textbf{DS-AGC}\\ 
\midrule
\multirow{3}{*}{\textbf{$Label_3$}}&
$2:1$ & 44.20$\pm$00.83 & 44.09$\pm$00.54 & 44.15$\pm$00.66 & 44.19$\pm$00.74 & \textbf{44.73$\pm$00.59} \\\hspace*{\fill}&
$1:1$ &	44.17$\pm$00.50 & 44.06$\pm$00.63 & 44.02$\pm$00.51 & 44.16$\pm$00.63 & \textbf{44.59$\pm$00.51} \\\hspace*{\fill}&
$1:2$ &	44.08$\pm$00.60 & 44.03$\pm$00.74 & 44.14$\pm$00.66 & 44.13$\pm$00.64 & \textbf{44.46$\pm$00.49} \\
\midrule
\multicolumn{2}{l}{Average Performance}   & 44.15$\pm$00.66 & 44.06$\pm$00.64 & 44.11$\pm$00.62 &  44.16$\pm$00.67 & \textbf{44.59$\pm$00.54} \\
\midrule
\specialrule{0em}{1.5pt}{1.5pt}
\midrule
\multirow{3}{*}{\textbf{$Label_9$}}&
$2:1$ & 14.43$\pm$00.16 & 14.41$\pm$00.16 & 14.40$\pm$00.11 & 14.41$\pm$00.07 & \textbf{15.02$\pm$00.24} \\\hspace*{\fill}&
$1:1$ &	14.41$\pm$00.14 & 14.36$\pm$00.11 & 14.41$\pm$00.14 & 14.40$\pm$00.15 & \textbf{15.13$\pm$00.13} \\\hspace*{\fill}&
$1:2$ &	14.38$\pm$00.19 & 14.40$\pm$00.18 & 14.40$\pm$00.05 & 14.39$\pm$00.10 & \textbf{15.00$\pm$00.33} \\
\midrule
\multicolumn{2}{l}{Average Performance} & 14.41$\pm$00.16 & 14.39$\pm$00.16 & 14.40$\pm$00.11 &  14.40$\pm$00.11 & \textbf{15.05$\pm$00.25} \\
\bottomrule
\end{tabular*}}
\end{center}
\end{table*}

\section{Discussion and Conclusion}
In order to thoroughly assess the performance of the proposed model, we conduct a series of experiments to evaluate the contribution of each module in the proposed model and also explore the effect of various hyperparameter settings.

\subsection{Ablation Study}

\textcolor{black}{To assess the contribution of each component in the proposed model, a series of ablation studies are conducted. Specifically, we systematically removed different components to observe their impact on the overall performance. Table \ref{tab:ablation} reports the ablation results using cross-subject leave-one-subject-out cross-validation under $N=2$. Firstly, we remove the discriminator from the model and find that the performance significantly drops when neglecting the distributional differences among domains. The model's performance decrease from $87.37\pm06.19$ to $78.71\pm06.52$ on SEED, from $66.00\pm07.93$ to $64.03\pm08.60$ on SEED-IV, and from $59.40\pm09.99$ to $53.32\pm10.95$ on SEED-V. 
\textcolor{black}{Secondly, we evaluate the benefits of considering three domains (labeled source domain, unlabeled source domain, and target domain) and compare the proposed semi-supervised multi-domain adaptation method with traditional domain adaptation methods, which treat the labeled and unlabeled source domains as a single domain. The results demonstrate that considering feature alignment between both labeled and unlabeled source data achieves higher performance, leading to improvements of 2.44\%, 1.50\%, and 5.06\% on the SEED, SEED-IV, and SEED-V databases, respectively.}
Thirdly, we evaluate the model without contrastive learning, directly extracting structural features from the constructed GCN. The model performance falls from $87.37\pm06.19$ to $86.48\pm05.07$ for SEED, from $66.00\pm07.93$ to $62.98\pm07.29$ for SEED-IV, and from $59.40\pm09.99$ to $57.33\pm11.17$ for SEED-V. It shows the significance of contrastive learning in improving feature discrimination and enhancing overall model performance. Fourthly, the importance of attentive feature fusion is evaluated. When the classifier is developed based on a simple concatenation of the extracted non-structural and structural features, a performance decline is observed across all three databases: from $87.37\pm06.19$ to $84.23\pm05.47$ for SEED, from $66.00\pm07.93$ to $64.39\pm07.83$ for SEED-IV, and from $59.40\pm09.99$ to $57.25\pm13.95$ for SEED-V. Lastly, the effect of attentive sample selection is examined. Equal weights ($\operatorname{Sim}(r_s^b)=1$) in Eq. \ref{Eq:class} are assigned to all samples. The results show that considering sample importance could be beneficial to the performance, bringing an average improvement in model performance by 0.45\%, 0.57\% and 0.97\% on SEED, SEED-IV, and SEED-V databases, respectively.} The above results provide a detailed breakdown of the performance of the model with and without specific components, indicating that each component in the proposed model plays an important role in improving the overall performance and addressing the challenges of emotion recognition in real-world scenarios with incomplete label information.

\begin{table*}[]
\begin{center}
\color{black}
\caption{\textcolor{black}{The ablation results using cross-subject leave-one-subject-out cross-validation under $N = 2$.}}
\label{tab:ablation}
\scalebox{1}{
\begin{tabular*}{\hsize}{@{}@{\extracolsep{\fill}}lccc@{}}
\toprule
Methods   & SEED & SEED-IV & SEED-V  \\ 
\midrule
Without discriminator	            &78.71$\pm$06.52 &64.03$\pm$08.60 &53.32$\pm$10.95\\
With traditional domain adaptation  &84.93$\pm$07.06 &64.50$\pm$08.76 &54.34$\pm$13.65\\
Without contrastive learning	    &86.48$\pm$05.07 &62.98$\pm$07.29 &57.33$\pm$11.17\\
Without attentive feature fusion    &84.23$\pm$05.47 &64.39$\pm$07.83 &57.25$\pm$13.95\\
Without attentive sample selection  &86.92$\pm$05.87 &65.43$\pm$08.24 &58.43$\pm$11.77\\
\midrule
\textbf{DS-AGC}  & \textbf{87.37$\pm$06.19} & \textbf{66.00$\pm$07.93} & \textbf{59.40$\pm$09.99}\\
\bottomrule
\end{tabular*}}
\end{center}
\end{table*}

\textcolor{black}{Additionally, we further assess the contribution of the dual-stream design and the incorporation of unlabeled data in the model learning process. To evaluate the dual-stream design, we analyze the model's performance with each stream separately under various $N$ values. As shown in Fig. \ref{fig:ablation}, whether using a single structural stream (pink line with circle marker) or a single non- structural stream (green line with square marker), their performance across all settings ($N=1:13$) is inferior to that of the complete dual-stream design (red line with triangle marker).}

\begin{figure}
\begin{center}
\includegraphics[width=0.5\textwidth]{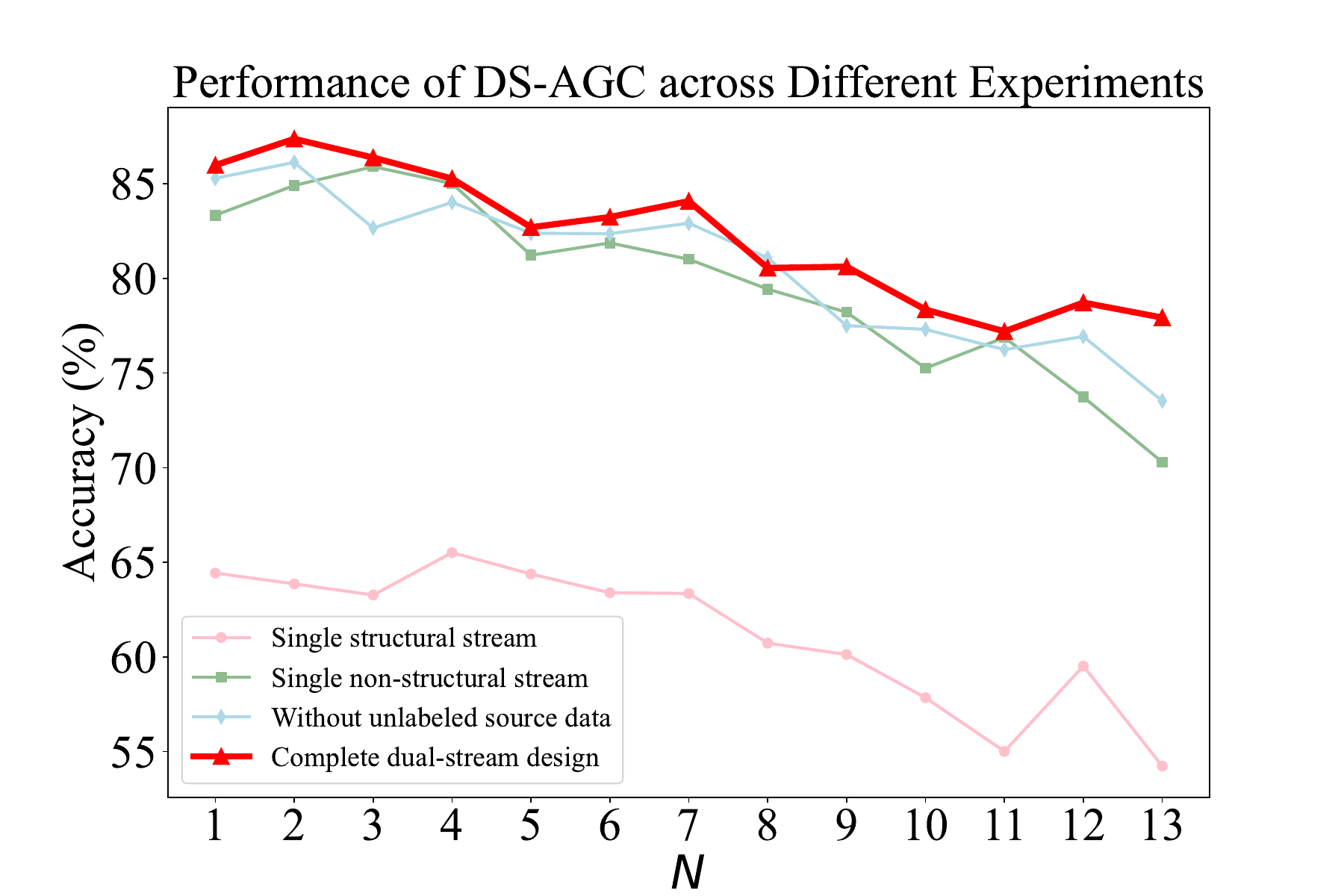}
\end{center}
\caption{\textcolor{black}{Experimental results under different settings on the SEED databases. Red line with triangle marker: the proposed DS-AGC; pink line with circle marker: single structural stream; green line with square marker: single non- structural stream; blue line with diamond marker: without unlabeled source data.}}
\label{fig:ablation}
\end{figure}

\textcolor{black}{In evaluating the contribution of the unlabeled source data in modeling, we compare the model's performance both with and without the inclusion of the unlabeled source data. As shown in Fig. \ref{fig:ablation}, the model performance without unlabeled source data (blue line with diamond marker) shows poorer results compared to the performance with unlabeled source data. It demonstrates that including the unlabeled source data plays an important role in the model learning process. Additionally, we conduct supervised learning for comparison, where all the source data are labeled. Following the same cross-validation protocol, the supervised learning results on SEED, SEED-IV, and SEED-V are $87.72\pm06.54$, $67.95\pm07.84$, and $60.27\pm11.92$, respectively. These results indicate that the proposed DS-AGC model can achieve relatively stable results similar to supervised learning in a semi-supervised manner, even when labeled data is limited.}

\subsection{Model performance on unseen target data}
\textcolor{black}{We further explore the model stability on completely unseen target data. For this purpose, we divide the target domain data into two parts: one as a validation set (visible in the model learning phase) and the other as an unseen test set (completely invisible in the model learning phase). The ratio of the validation set to the unseen set is controlled by the parameter $Z$. When $Z=0$, it indicates that all the target domain data are assigned as the unseen test set. When $Z=15$ (SEED and SEED-V) or $Z=24$ (SEED-IV), it indicates that all the target domain data are assigned as the validation set, as utilized in Section \ref{sec:experiment}. The corresponding results under various $Z$ values on the SEED, SEED-IV, and SEED-V databases are reported in Table \ref{tab:unseenTargetSEED}, \ref{tab:unseenTargetSEEDIV}, and \ref{tab:unseenTargetSEEDV}, respectively. It shows that an increase in the $Z$ value, which brings more information during the model learning phase, could be helpful in improving the model performance. Even in the extreme case with $Z=0$ (all the target domain data is completely unseen during model learning), it still manages to exhibit acceptable performance. It demonstrates the model's good generalization capability and accuracy in predicting unknown data, highlighting its ability to handle incomplete label problems in real-world application scenarios.}

\begin{table}[]
\centering
\color{black}
\caption{Model performance on unseen SEED data, where the total number of trials of each subject is 15.}
\label{tab:unseenTargetSEED}
\setlength{\tabcolsep}{5mm}
\scalebox{1}{
\begin{tabular}{cccc} 
\toprule
Validation Set $\mathbb{T}_{v}$  & Test Set $\mathbb{T}_{t}$  & \multirow{2}{*}{$P_{acc}$} \\
($Z$ trials) & (($15-Z$) trials) & \\
\midrule
0 & 15 & 79.17$\pm$04.79 \\
3 & 12 & 78.85$\pm$05.31  \\ 
6 & 9  & 79.33$\pm$07.72  \\ 
9 & 6  & 82.15$\pm$06.57  \\ 
12 & 3 & 82.44$\pm$14.72  \\ 
\textbf{15} & \textbf{15} & \textbf{87.37}$\pm$\textbf{06.19}  \\ 
\bottomrule
\end{tabular}
}
\end{table}

\begin{table}[]
\centering
\color{black}
\caption{Model performance on unseen SEED-IV data, where the total number of trials of each subject is 24.}
\label{tab:unseenTargetSEEDIV}
\setlength{\tabcolsep}{5mm}
\scalebox{1}{
\begin{tabular}{cccc} 
\toprule
Validation Set $\mathbb{T}_{v}$  & Test Set $\mathbb{T}_{t}$  & \multirow{2}{*}{$P_{acc}$} \\
($Z$ trials) & (($24-Z$) trials) & \\
\midrule
0 & 24   & 63.02$\pm$08.63 \\
4 & 20   & 61.44$\pm$06.94  \\ 
8 & 16   & 62.49$\pm$08.72  \\ 
12 & 12  & 62.21$\pm$08.07  \\ 
16 & 8   & 63.47$\pm$12.09  \\ 
20 & 4   & 61.16$\pm$13.45  \\ 
\textbf{24} & \textbf{24}   & \textbf{66.00}$\pm$\textbf{07.93} \\ 
\bottomrule
\end{tabular}
}
\end{table}

\begin{table}[]
\centering
\color{black}
\caption{Model performance on unseen SEED-V data, where the total number of trials of each subject is 15.}
\label{tab:unseenTargetSEEDV}
\setlength{\tabcolsep}{5mm}
\scalebox{1}{
\begin{tabular}{cccc} 
\toprule
Validation Set $\mathbb{T}_{v}$  & Test Set $\mathbb{T}_{t}$  & \multirow{2}{*}{$P_{acc}$} \\
($Z$ trials) & (($15-Z$) trials) & \\
\midrule
0 & 15   & 54.07$\pm$10.82 \\
5 & 10   & 48.47$\pm$08.22  \\ 
10 & 5   & 56.03$\pm$13.58  \\ 
\textbf{15} & \textbf{15}   & \textbf{59.40}$\pm$\textbf{09.99} \\ 
\bottomrule
\end{tabular}
}
\end{table}

\subsection{The effect of the hyperparameter $E_t$}
\textcolor{black}{During the training process, the unlabeled source domain $\mathbb{U}$ is incorporated when the model has undergone an initial warm-up period of $E_t$ epochs. We evaluate how different $E_t$ values influence model performance, with $E_t$ ranging from \textcolor{black}{0} (start of training) to 100 (maximum number of epochs). As shown in Fig. \ref{fig:Etvalue}, optimal results are obtained when $\mathbb{U}$ is incorporated once the model has achieved a certain degree of stability, rather than from the very beginning ($E_t=0$). Introducing $\mathbb{U}$ at the onset can inject noise, potentially disrupting the model's initial learning phase.}

\begin{figure}
\begin{center}
\includegraphics[width=0.5\textwidth]{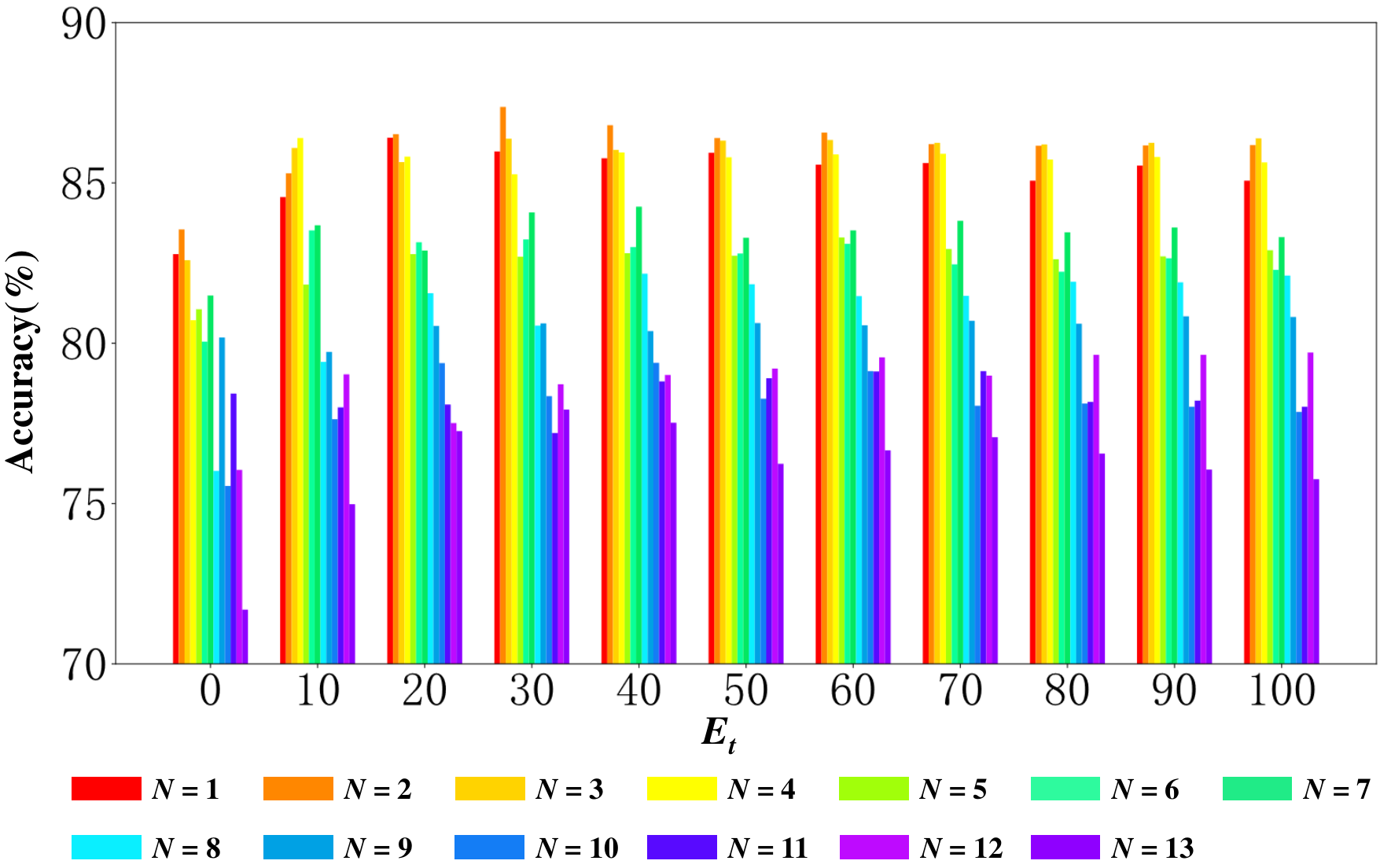}
\end{center}
\caption{\textcolor{black}{Model performance under various $E_t$ values on the SEED database.}}
\label{fig:Etvalue}
\end{figure}

\subsection{Data visualization}
To delve deeper into the learning process, we employ the t-distributed stochastic neighbor embedding (t-SNE) algorithm \cite{van2008visualizing} to visually compare the acquired feature representation at different stages. This analysis allows us to gain valuable insights into the model's learning dynamics. Specifically, we visualize the fused features obtained through $MHA(\cdot)$ (Fig. \ref{fig:MHAfeature}) and the final classification results (Fig. \ref{fig:classifyResults}) at various learning stages: before training, at the 30th training epoch, and in the final trained model. By examining these visualizations at different learning stages, we obtain vivid depictions of the evolution and enhancement of both the feature representation and the classification performance. These visualizations showcase a notable expansion of inter-class separability, meaning that the distinctions between different classes become more pronounced. Simultaneously, the intra-class variability is minimized, resulting in tighter clustering of samples belonging to the same class. Through this visual examination, we observe a clear trend of the model's ability to discriminate between different classes, with the learned feature representation becoming increasingly distinct and discriminative. The reduction in intra-class variability ensures that samples within the same class are closer to each other, reinforcing the model's ability to accurately classify them. This visualization evidence underscores the model's capability to learn meaningful and discriminative features, enabling it to make refined distinctions between classes and achieve enhanced classification performance throughout the learning process.

\begin{figure*}
\begin{center}
\includegraphics[width=\textwidth]{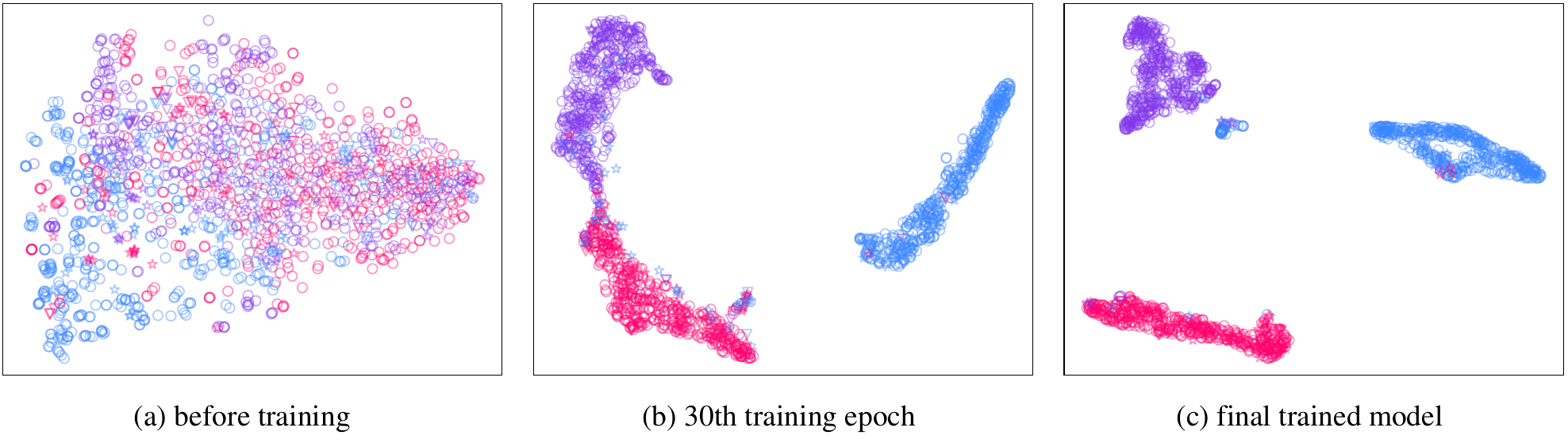}
\end{center}
\caption{A visualization of the obtained informative feature $\{\xi_{1}, \ldots, \xi_{\dot{m}+\dot{n}}\}$ by $MHA(\cdot)$ at three different stages: (a) before training, (b) at the 30th training epoch, and (c) in the final trained model. In this visualization, the circle, asterisk, and triangle represent the labeled source domain ($\mathbb{S}$), unlabeled source domain ($\mathbb{U}$), and the unknown target domain ($\mathbb{T}$), respectively. The \textcolor{black}{red, purple, and blue} colors correspond to negative, neutral, and positive emotions, respectively.}
\label{fig:MHAfeature}
\end{figure*}

\begin{figure*}
\begin{center}
\includegraphics[width=\textwidth]{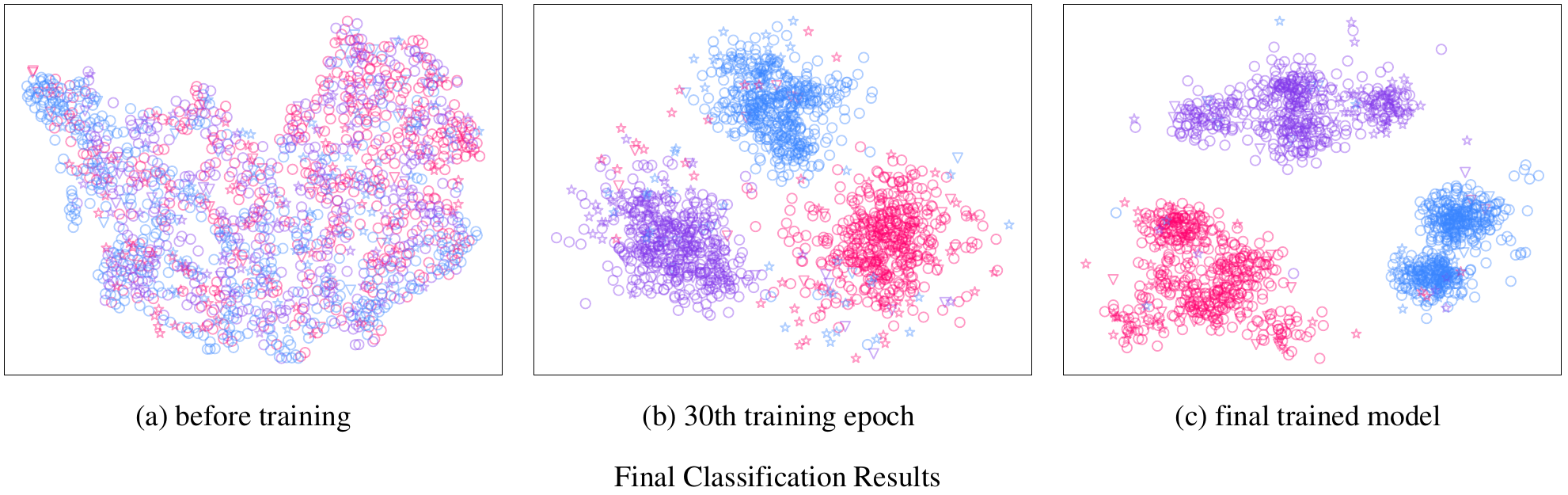}
\end{center}
\caption{A visualization of the classification results at three different stages: (a) before training, (b) at the 30th training epoch, and (c) in the final trained model. In this visualization, the circle, asterisk, and triangle represent the labeled source domain ($\mathbb{S}$), unlabeled source domain ($\mathbb{U}$), and the unknown target domain ($\mathbb{T}$), respectively. The \textcolor{black}{red, purple, and blue} colors correspond to negative, neutral, and positive emotions, respectively.}
\label{fig:classifyResults}
\end{figure*}

\color{black}
\subsection{Conclusion}
This paper proposes a novel semi-supervised dual-stream self-attentive adversarial graph contrastive learning model (DS-AGC) for cross-subject EEG-based emotion recognition. Through an intelligent dual-stream framework, both non-structural and structural information from EEG signals are well established and further fused through a self-attentive mechanism. Based on comprehensive experimental validations on the well-known databases, the proposed DS-AGC shows the ability to effectively leverage both labeled and unlabeled data and also makes it a promising solution for situations where labeled data is scarce or expensive to obtain.

\section{Acknowledgments}
This work was supported in part by the National Natural Science Foundation of China under Grant 62276169, 62071310, and 82272114, in part by the Medical-Engineering Interdisciplinary Research Foundation of Shenzhen University under Grant 2024YG008, in part by the Shenzhen University-Lingnan University Joint Research Programme, in part by the Shenzhen-Hong Kong Institute of Brain Science-Shenzhen Fundamental Research Institutions under Grant 2022SHIBS0003, in part by Shenzhen Science and Technology Research and Development Fund for Sustainable Development Project under Grant KCXFZ20201221173613036.


%




\ifCLASSOPTIONcaptionsoff
  \newpage
\fi

\bibliographystyle{IEEEtran}
\bibliography{references}

\begin{IEEEbiography}[{\includegraphics[width=1in,height=1.25in,clip,keepaspectratio]{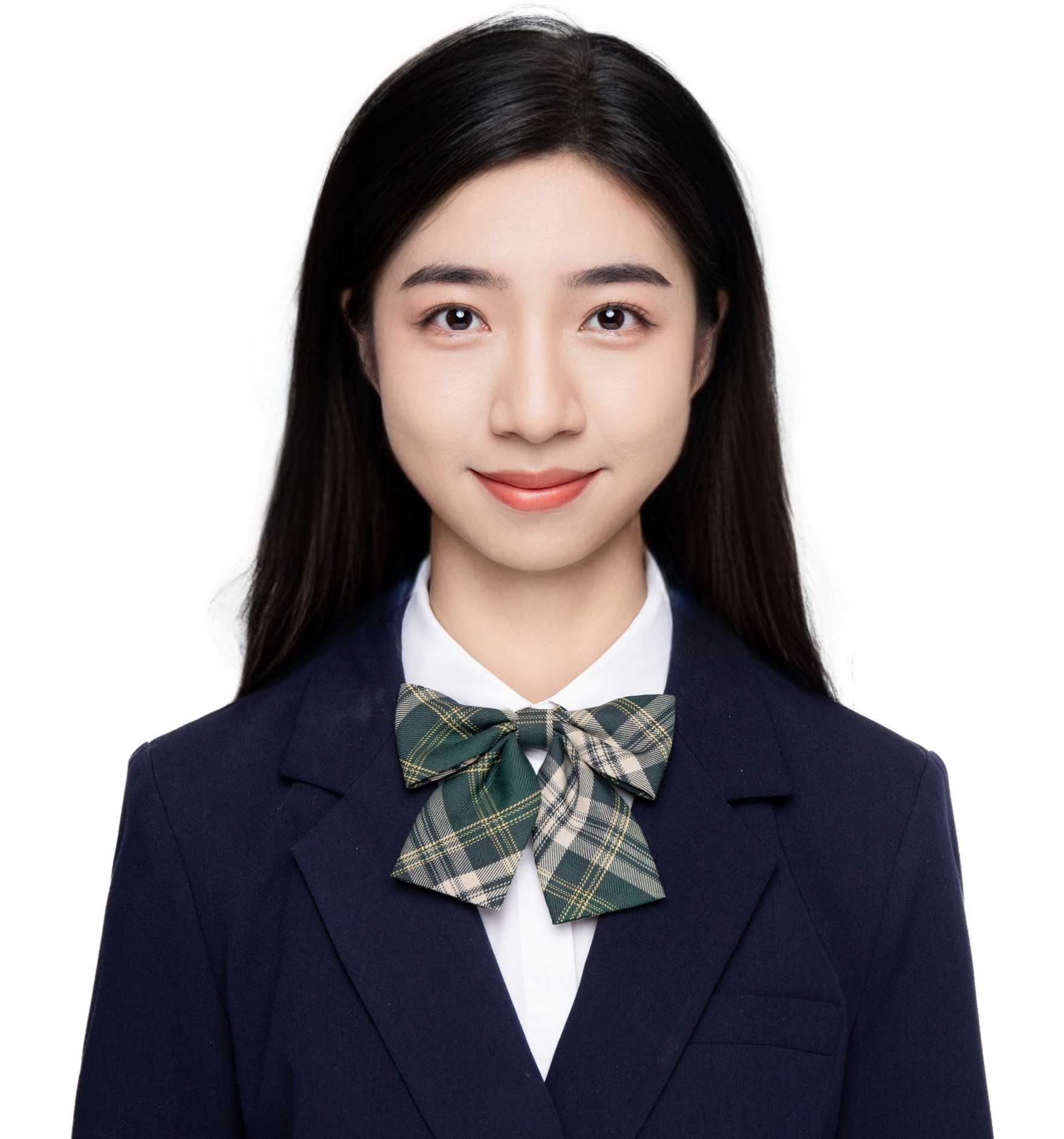}}]{Weishan Ye}
\textcolor{black}{
 is a master's student in the Department of Biomedical Engineering at Shenzhen University, China. Her research interests include affective computing, semi-supervised learning, and transfer learning.
 }
\vspace{-10mm}
\end{IEEEbiography}

\begin{IEEEbiography}[{\includegraphics[width=1in,height=1.25in,clip,keepaspectratio]{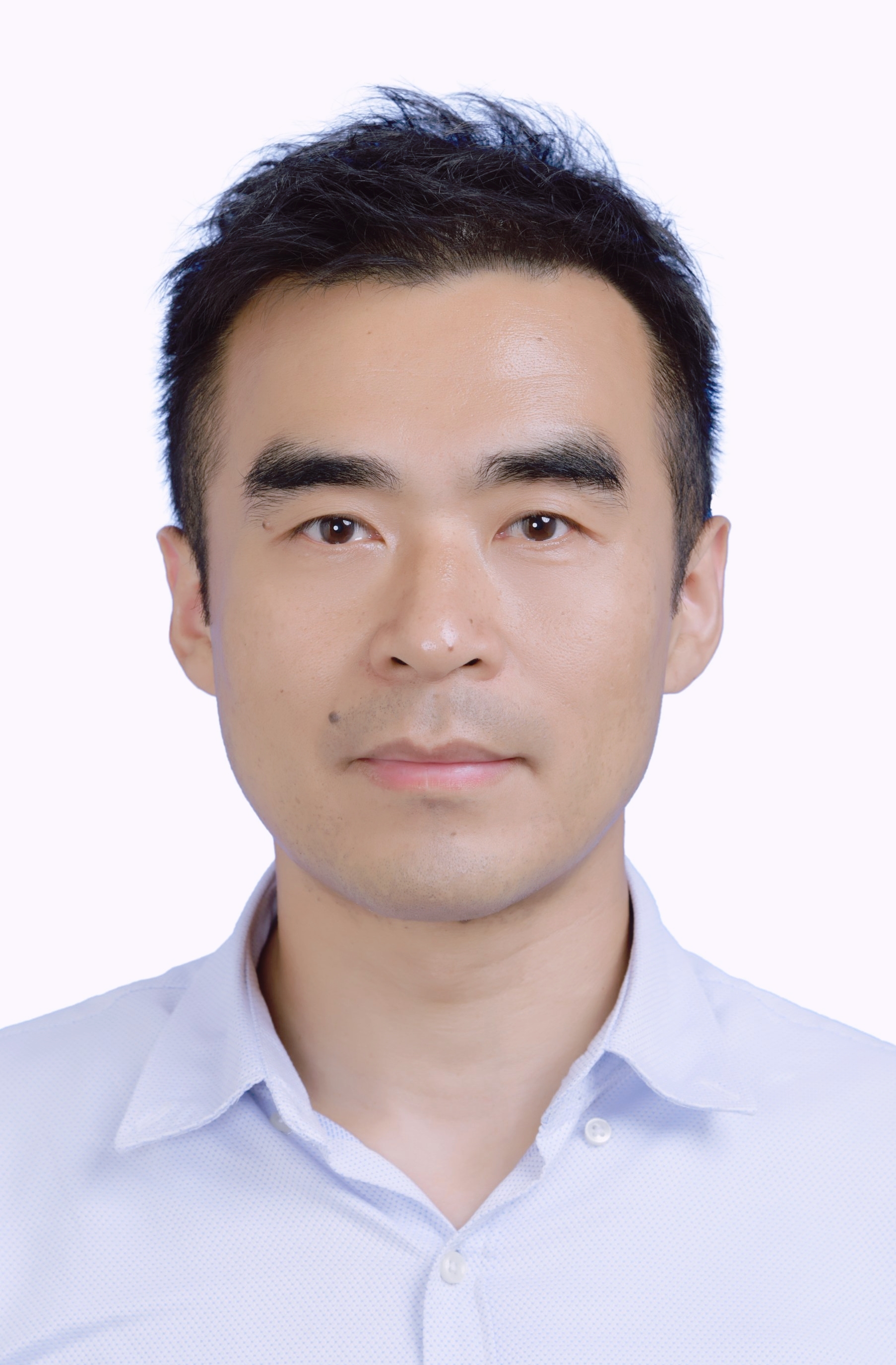}}]{Zhiguo Zhang}
received his B.Eng. degree from Tianjin University in 2000, his M.Phil. degree from the University of Science and Technology of China in 2003, and his Ph.D. degree from the University of Hong Kong in 2008. He is now a Professor with the School of Computer Science and Technology, Harbin Institute of Technology, Shenzhen, China. He has authored 130 journal papers. His research interests include biomedical signal processing, neural engineering, and brain-inspired computation.
\vspace{-10mm}
\end{IEEEbiography}

\begin{IEEEbiography}[{\includegraphics[width=1in,height=1.25in,clip,keepaspectratio]{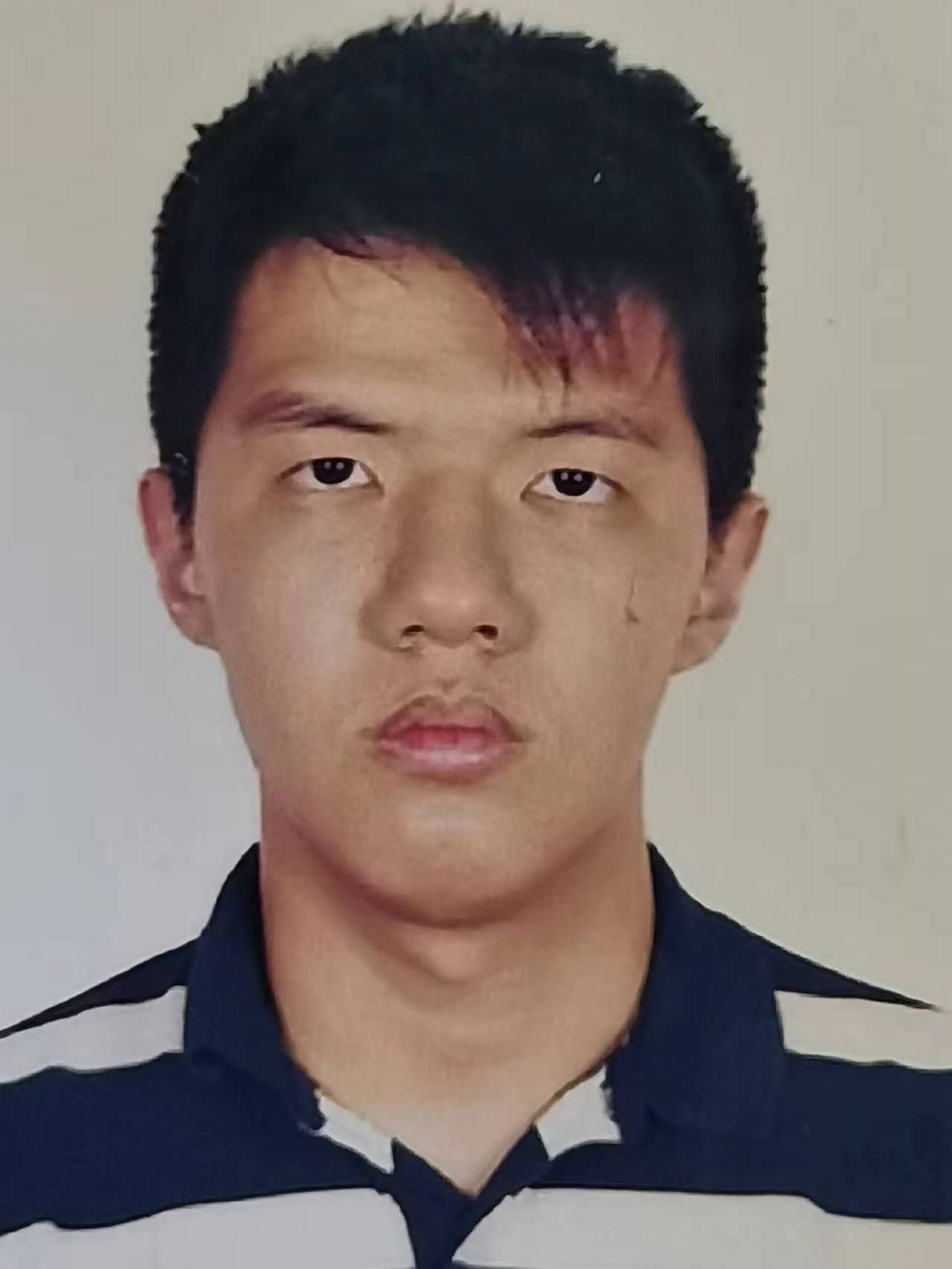}}]{Fei Teng}
is a bachelor student in the Department of Biomedical Engineering, Shenzhen University, China. His research interests include transfer learning and affective brain-computer interface.
\vspace{-10mm}
\end{IEEEbiography}

\begin{IEEEbiography}[{\includegraphics[width=1in,height=1.25in,clip,keepaspectratio]{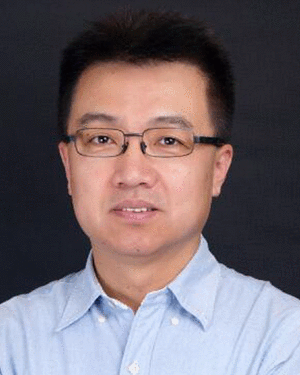}}]{Min Zhang}
received the bachelor's and Ph.D. degrees in computer science from the Harbin Institute of Technology, Harbin, China, in 1991 and 1997, respectively. He is currently a Professor with with the School of Computer Science and Technology, Harbin Institute of Technology, Shenzhen, China. He has authored 180 papers in leading journals and conferences. His research interests include machine translation, natural language processing, and artificial intelligence. He is the Vice President of COLIPS, Steering Committee Member of PACLIC, Executive Member of AFNLP, and Member of ACL.
\vspace{-10mm}
\end{IEEEbiography}

\begin{IEEEbiography}[{\includegraphics[width=1in,height=1.25in,clip,keepaspectratio]{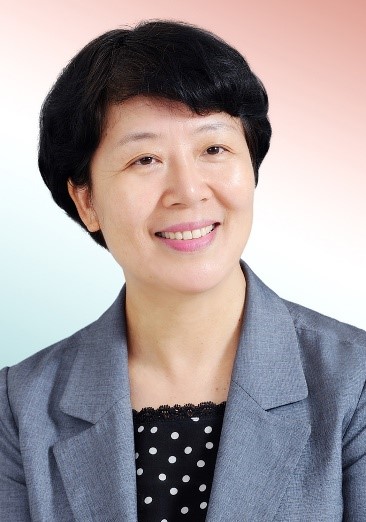}}]{Jianhong Wang}
received her MSHA degree from Nankai University-Flinders University in Australia, in 2008. From 2016 to 2019, she was vice president of Shenzhen People’s Hospital. From 2019 to now, she has served as Secretary of the Party Committee of Shenzhen Kangning Hospital. She is also currently a master tutor of Southern University of Science and technology. She has been engaged in hospital management for nearly 20 years, accumulated a lot of hospital management experience and achieved many achievements. Her current research interests include hospital management, public health, and psychologic medicine.
\vspace{-10mm}
\end{IEEEbiography}

\begin{IEEEbiography}[{\includegraphics[width=1in,height=1.25in,clip,keepaspectratio]{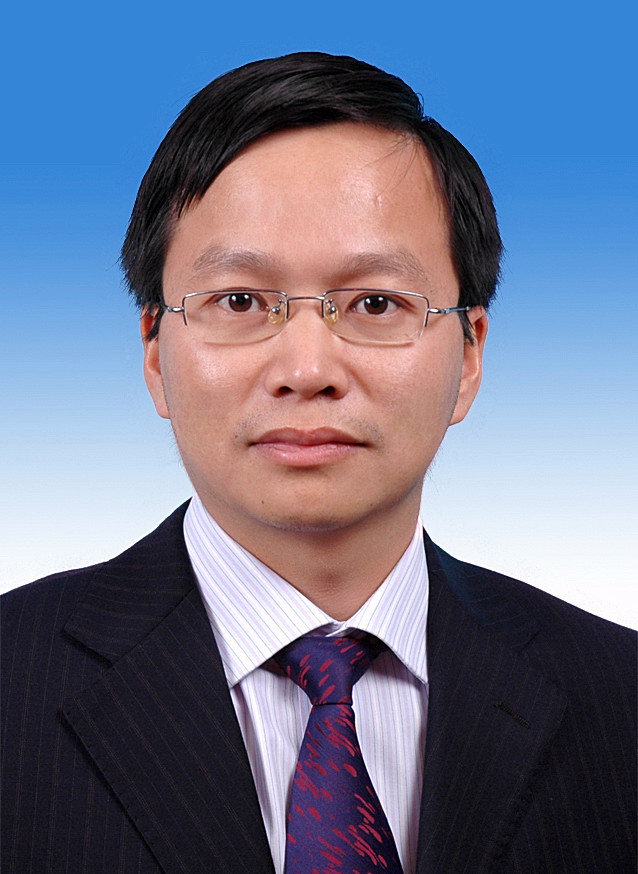}}]{Dong Ni}
received his Bachelor and Master degrees in Biomedical Engineering from Southeast University, Nanjing, China, in 2000 and 2003, respectively. He received a PhD degree in computer science and engineering from the Chinese University of Hong Kong, Hong Kong, China in 2009. From 2009 to 2010, he was a postdoctoral fellow with the School of Medicine, University of North Carolina at Chapel Hill, USA. He has joined Shenzhen University since 2010. In 2013, he became Associate Professor in Biomedical Engineering at Shenzhen University. Due to outstanding achievements, he was promoted to Professor in Biomedical Engineering of Shenzhen University in 2017. He has been the Vice Dean of the School of Biomedical Engineering of Shenzhen University since 2016. Dong founded the Medical Ultrasound Image Computing (MUSIC) Laboratory in Shenzhen University in 2018. His research interests include ultrasound image analysis, image guided surgery, and machine/deep learning. He has published over 200 peer-reviewed journal/conference articles. He is also committed to translate multiple research findings into products to help increase the efficiency and standardization of ultrasound diagnostics. Dong is a long-standing member of the MICCAI Society. He is one of the founders of the Medical Image Computing Seminar (MICS) in China. He served as the chairs of both Local Organization committee and Finance committee in MICCAI 2019 and endeavored to make this conference the most successful in the history of MICCAI. In January 2020, he was elected as Member of MICCAI Board.
\vspace{-10mm}
\end{IEEEbiography}

\begin{IEEEbiography}[{\includegraphics[width=1in,height=1.25in,clip,keepaspectratio]{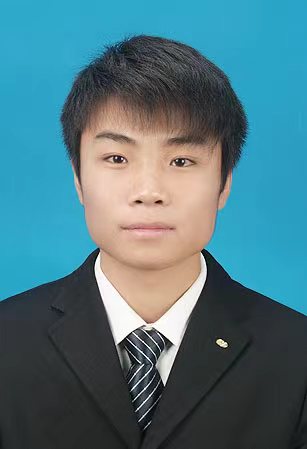}}]{Fali Li}
received his Ph.D degree from University of Electronic Science and Technology of China in 2020. He is now an associate researcher in the School of Life Science and Technology, Center for Information in Medicine, University of Electronic Science and Technology of China, China. His research interests include brain-computer interface, bio-informatics, and neural engineering.
\vspace{-10mm}
\end{IEEEbiography}

\begin{IEEEbiography}[{\includegraphics[width=1in,height=1.25in,clip,keepaspectratio]{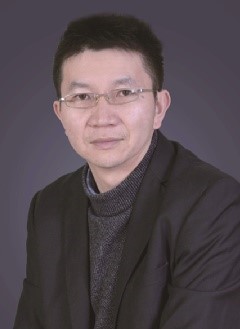}}]{Peng Xu}
received the Ph.D. degree in Biomedical Engineering from the School of Life Science and Technology, University of Electronic Science and Technology of China in 2006. He was a postdoctoral researcher at UCLA from 2007-02 to 2009-04. He is currently a full professor at the School of Life Science and Technology, University of Electronic Science and Technology of China, China. His research interests include brain–computer interface, brain inspired intelligence, machine learning, and brain network analysis, etc.
\vspace{-10mm}
\end{IEEEbiography}

\begin{IEEEbiography}[{\includegraphics[width=1in,height=1.25in,clip,keepaspectratio]{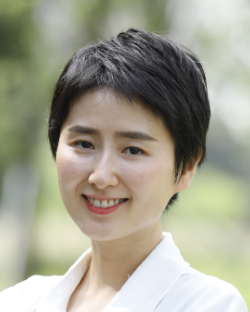}}]{Zhen Liang}
received her Ph.D. degree from The Hong Kong Polytechnic University, Hong Kong, in 2013. From 2012 to 2017, she was an algorithm development scientist at NeuroSky, Inc., Hong Kong. From 2018 to 2019, she was a specially‐appointed assistant professor of Graduate School of Informatics, Kyoto University, Japan. She is currently an associate professor in the School of Biomedical Engineering, Health Science Center, Shenzhen University, China. Her current research interests include brain encoding and decoding systems, affective computing, visual attention, and neural engineering.
\vspace{150mm}
\end{IEEEbiography}












\end{document}